\documentclass{article}
\usepackage{amsmath}
\usepackage{amssymb}
\usepackage{graphicx} 

\begin{document}

\title{A simple derivation of the Tracy-Widom distribution of the 
maximal eigenvalue of a Gaussian unitary 
random matrix}
\author{Celine Nadal and Satya N. Majumdar \\
\\
Univ. Paris-Sud, CNRS, LPTMS, UMR 8626, Orsay F-01405, France}

\date{}

\maketitle

\begin{abstract}
 
In this paper, we first briefly review some recent results on the distribution
of the maximal eigenvalue of a $(N\times N)$ random matrix drawn from Gaussian ensembles.
Next we focus on the Gaussian Unitary Ensemble (GUE) and
by suitably adapting a method of orthogonal polynomials developed
by Gross and Matytsin in the context of Yang-Mills theory in two dimensions, 
we provide a rather simple derivation of the Tracy-Widom law for GUE. 
Our derivation is based on the elementary asymptotic scaling analysis
of a pair of coupled nonlinear recursion relations. 
As an added bonus, this method also allows us to compute the precise subleading terms describing
the right large deviation tail of the maximal eigenvalue distribution. In the Yang-Mills
language, these subleading terms correspond to non-perturbative (in $1/N$ expansion) corrections to
the two-dimensional partition function in the so called `weak' coupling regime.

\end{abstract}

\section{Introduction}

Quite a long time ago, Wigner~\cite{wigner} introduced random matrices in the context of nuclear 
physics. He suggested that the highly-excited energy levels of complex nuclei can locally be well 
represented by the eigenvalues of a large random matrix. A big nucleus is a rather complex system
composed of many strongly interacting quantum particles and it is practically impossible to 
describe its spectral properties via first principle calculations. The idea of Wigner was 
to model the spectral properties of the complex Hamiltonian of such a big nucleus by
those of a large random matrix preserving the same symmetry. This was a very successful
approach in nuclear physics.
Since then, 
the random matrix theory (RMT) has gone beyond nuclear physics and has found a wide number 
of applications in various fields of physics and 
mathematics including quantum chaos, disordered systems, string theory and even number 
theory~\cite{mehta}. A case of special interest is the one of Gaussian 
random matrices (originally introduced by 
Wigner himself) where the entries of the matrix are Gaussian random variables.

Depending on the symmetry of the problem, Dyson distinguished three 
classes
for the matrix $X$~\cite{Dyson}:
\newline
 $\diamond$  the Gaussian Orthogonal Ensemble (GOE) :  $X$ is real symmetric.
\newline
$\diamond$ the Gaussian Unitary Ensemble (GUE) : $X$ is complex Hermitian.
\newline
$\diamond$ the Gaussian Symplectic Ensemble (GSE) : $X$ is quaternionic Hermitian.


Let us write $X^{\dagger}$ the adjoint of $X$, i.e. the transpose of $X$ for the GOE,
the complex conjugate transpose for the GUE and the quaternionic conjugate transpose
for the GSE.  
A Gaussian random matrix is a $N\times N$  self-adjoint matrix $X$, i.e. $X^{\dagger}=X$
distributed according to the law
\begin{equation}
\label{entrydist}
 \mathcal{P}(X)\propto e^{-\frac{\beta}{2}{\rm Tr}\left(X^2 \right)}\;\;\;\;\;\; {\rm with}\;\;
\beta=\left\{ \begin{array}{lll} 1&\textrm{ for GOE }\\
                                 2&\textrm{ for GUE }\\   
                                 4&\textrm{ for GSE }
              \end{array}\right.
\end{equation}
where, for convenience, we have chosen the prefactor $\beta$ of the ${\rm Tr}(X^2)$ 
to be $\beta=1$ for the GOE, $\beta=2$ for the GUE  and $\beta=4$ for the GSE.
For instance, for  the GUE we have $\beta=2$
and
 $\mathcal{P}(X)\propto e^{-{\rm Tr}\left(X^2 \right)}
\propto e^{-\sum_{i,j} \left|X_{i,j}\right|^{2}}$ as $X^2=X^{\dagger} X= \sum_{i,j} |X_{i,j}|^2$. 
This means that $X$ is a $N\times N$  complex Hermitian matrix with entries
 ${\rm Re} X_{i,j}$ and ${\rm Im} X_{i,j}$ for $i<j$ that are independent (real) random variables
distributed according to the same centered Gaussian law with variance $1/4$ 
and the $X_{i,i}$ are (real) independent Gaussian variables with mean $0$ and 
variance $1/2$. In case of GSE, there are $2N$ eigenvalues, each 
of them two-fold degenerate and {\rm Tr} 
in \eqref{entrydist} for $\beta=4$ is defined
so that only one of the two fold degenerate eigenvalues in $X$ is counted.
\\

Self-adjoint matrices can be diagonalized and have real eigenvalues.
The joint distribution of eigenvalues of the Gaussian ensemble is 
well known~\cite{PR,mehta}  
\begin{equation}\label{eq:jpdfEV}
 \mathcal{P}(\lambda_1,...\lambda_N)=B_N\; e^{-\frac{\beta}{2} \sum_{i=1}^N \lambda_i^2}\;
\prod_{j<k} \left|\lambda_j-\lambda_k\right|^{\beta} 
\end{equation}
where $B_N$ is a normalization constant such that $\int \left( \prod_i d\lambda_i\right) \mathcal{P}(\lambda_1,...,\lambda_N)=1$
(it depends on $\beta$)
and the power $\beta$ of the Vandermonde term is called the Dyson index $\beta=1, 2 \: {\rm or} \: 4$ 
depending on the ensemble (resp. GOE, GUE or GSE). Note that we have chosen the prefactor
of ${\rm Tr}(X^2)$ term in \eqref{entrydist} to be the same as the Dyson index $\beta$
just for convenience. This prefactor is not very 
important as it can be absorbed by rescaling
the matrix entries by a constant factor. In contrast,
the value of the Dyson index $\beta=1$, $2$ or $4$, characterizing the 
power of the Vandermonde term, plays a crucial role.
The normalization constant $B_N$ can be computed using Selberg's integral~\cite{mehta}:
$B_N={\beta}^{\frac{N}{2}+\beta \frac{N (N-1)}{4}}\, (2 \pi)^{-\frac{N}{2}}\,
\Gamma(1+\beta/2)^N/\left[\prod_{j=1}^N \Gamma(1+j \beta/2)\right]$.

Because of the presence of the Vandermonde determinant  $\prod_{j<k} \left(\lambda_j-\lambda_k\right)$
in Eq. \eqref{eq:jpdfEV}, the eigenvalues
are strongly correlated random variables, they repel each other. 
In this paper, our focus is on the statistical properties of the extreme (maximal) eigenvalue
$\lambda_{\rm max}={\rm max}\left(\lambda_1, \lambda_2,\ldots,\lambda_N\right)$.
Had the Vandermonde term been not there in the joint distribution 
\eqref{eq:jpdfEV}, the
joint distribution would factorize 
and the
eigenvalues would thus be completely independent random variables, each with a Gaussian
distribution. For such independent and identically distributed random variables $\{\lambda_i\}$,
the extreme value statistics is well understood~\cite{Gumbel} and the distribution 
of the maximum, properly shifted and scaled,  belongs to one of the three universality classes Gumbel,
Frechet or Weibull (for large $N$) depending on the tail of the distribution of individual
$\lambda_i$'s.
However, in the case of random matrix theory, the eigenvalues $\lambda_i$'s are strongly correlated 
variables. For strongly correlated random variables there is no general theory for the
distribution of the maximum. In case of Gaussian random matrices, where the joint distribution \eqref{eq:jpdfEV}
is explicitly known, much progress has been made in understanding the distribution of $\lambda_{\rm max}$
following the seminal work by Tracy and Widom~\cite{TW1,TW2}. This then provides a very useful
solvable model for the extreme value distribution in a strongly correlated system and hence
is of special interest.   

\begin{figure}
 \includegraphics[width=12cm]{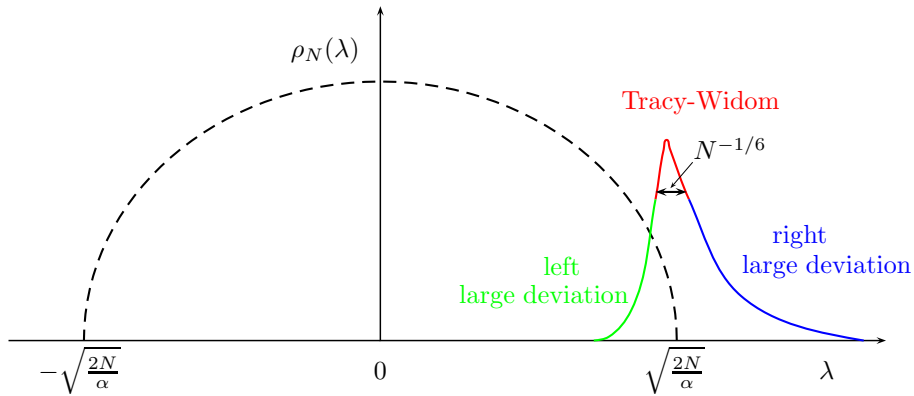}
\caption{Average density of the eigenvalues of  a Gaussian random matrix
$\rho_N(\lambda)$ as a function of $\lambda$ (blue dashed line). 
The density has a semi-circular shape (``Wigner semi-circle'')
and a finite support $[-\sqrt{\frac{2 N}{\alpha}},\sqrt{\frac{2 N}{\alpha}}]$.
The maximal eigenvalue has mean value $\langle \lambda_{\rm max}\rangle\approx \sqrt{\frac{2 N}{\alpha}}$
for large $N$ and its distribution close to the mean value, over a scale of $O(N^{-1/6})$ 
has the Tracy-Widom form (red solid line). However, over a scale $(\sqrt{N})$ the distribution
has large deviation tails shown by solid green (left large deviations) and solid blue (right large
deviations) lines.}
\label{fig:densTW}
\end{figure}

Let us first summarize some known properties of the random variable $\lambda_{\rm max}$.
Its average value can be easily obtained from the right edge of the well known
Wigner semi-circle describing the average density of eigenvalues.
For a Gaussian random matrix of large size
$N$, the average density of eigenvalues (normalized to unity) 
$\rho_N(\lambda)=\langle \frac{1}{N} \sum_i \delta(\lambda-\lambda_i)\rangle$
has a semi-circular shape on a finite support $[-\sqrt{2 N}, \sqrt{2N}]$ called the Wigner 
semi-circle~\cite{wigner}:
\begin{equation}\label{eq:averDens}
 \rho_N(\lambda)\approx \frac{1}{\sqrt{N}} \;g\left(\frac{\lambda}{\sqrt{N}} \right)\;\;{\rm with}
\;\; g(x)=\frac{1}{\pi}\sqrt{2-x^2}\;\;\;\textrm{ for large $N$}
\end{equation}
The quantity $\rho_N(\lambda)d\lambda$ represents the average fraction of eigenvalues that lie within the small interval $[\lambda, \lambda+d\lambda]$.
Therefore, Eq. \eqref{eq:averDens} means that the eigenvalues of a Gaussian random matrix lie on average 
within the finite interval $[-\sqrt{2 N}, \sqrt{2N}]$. 
Note also that one can rewrite, using the joint 
distribution in \eqref{eq:jpdfEV}
\begin{equation}
\rho_N(\lambda)=\langle \frac{1}{N} \sum_i 
\delta(\lambda-\lambda_i)\rangle=\int \mathcal{P}(\lambda,\lambda_2 
\dots, \lambda_N)\,d\lambda_2\dots d\lambda_N.
\label{marginal1}
\end{equation}
Hence the average density of states
$\rho_N(\lambda)$ can also be interpreted as the marginal
distribution of one of the eigenvalues (say the first one). 
Thus, the marginal distribution also 
has the shape of a semi-circle. 
Figure \ref{fig:densTW} shows the average density $\rho_N(\lambda)$
($\alpha=1$ here).
\\

It then follows that the average value of the maximal eigenvalue $\lambda_{\rm max}$ is  given for large $N$
by the upper bound of the density support: 
 \begin{equation}
  \langle \lambda_{\rm max}\rangle \approx \sqrt{2 N} \;\; \textrm{ for large $N$} 
 \end{equation}
However, $\lambda_{\rm max}$ fluctuates around this average value from one realization to another
and has a distribution around its mean value $\sqrt{2N}$ (see Fig. \ref{fig:densTW} with $\alpha=1$).
What is the full probability distribution of $\lambda_{\rm max}$? From the joint distribution
of eigenvalues in Eq. (\ref{eq:jpdfEV}), it is easy to write down formally the cumulative distribution
function (cdf)
of $\lambda_{\rm max}$ as a multiple integral
\begin{equation}
\mathbb{P}_N\left(\lambda_{\rm max} \leq t\right)= B_N\; \prod_{i=1}^N \int_{-\infty}^t d\lambda_i\:
\prod_{j<k} \left|\lambda_j-\lambda_k\right|^{\beta}\; e^{-\frac{\beta}{2} 
\sum_{i=1}^N \lambda_i^2}
\label{multint}
\end{equation}
which can be interpreted as a partition function of a Coulomb gas in presence of a
hard wall at the location $t$ (see the discussion in Section 2).
The question is how does $\mathbb{P}_N\left(\lambda_{\rm max} \leq t\right)$ behave for large $N$?
It turns out that the fluctuations of $\lambda_{\rm max}$ around its mean $\sqrt{2N}$
have two scales for large $N$.
While typical fluctuations scale as $N^{-1/6}$, large fluctuations scale as $N^{1/2}$ and
their probability distributions are described by different functional forms (see Fig. \ref{fig:densTW}
with $\alpha=1$). 

\vspace{0.2cm}

\noindent {\bf Typical fluctuations:} From an asymptotic analysis of the 
mutilple integral in Eq. (\ref{multint}), Forrester~\cite{Forrester1}, 
followed Tracy and Widom~\cite{TW1,TW2}
deduced that for large $N$, {\em small and typical} fluctuations of the maximal eigenvalue 
around its mean value $\sqrt{2N}$ are of order $O(N^{-1/6})$ and can be written as
\begin{equation}
 \lambda_{\rm max} \approx \sqrt{2 N}+a_{\beta} N^{-\frac{1}{6}} \, \chi
\end{equation}
where $a_{1,2}=1/\sqrt{2}$ (for GOE and GUE) and $a_4=2^{-7/6}$ (GSE)
and $\chi$ is a random variable characterizing the typical fluctuations.
Tracy and Widom~\cite{TW1,TW2} proved that for large $N$, the distribution 
of $\chi$
is independent of $N$: 
${\rm P}\left(\chi\leq x\right)=F_{\beta}(x)$. 
The function $F_{\beta}(x)$ depends
explicitly on $\beta$ and is called the Tracy-Widom distribution. For 
example, for $\beta=2$~\cite{TW1,TW2}, 
\begin{equation}
F_2(x)= \exp\left[-\int_x^{\infty} (z-x) q^2(z) dz\right]
\label{f2tw}
\end{equation}
where $q(z)$ satisfies the special case of $\alpha=0$ of the Painlev\'e II 
equation
\begin{equation}
q''(z) = 2 q^3(z) + z\,q(z)+ \alpha.
\label{painleve1}
\end{equation}
For $\alpha=0$, the solution only requires the right tail boundary 
condition for its unique specification: $q(z)\sim {\rm Ai}(z)$ as $z\to 
\infty$, where ${\rm Ai}(z)$ is the Airy function that
satisfies the differential equation ${\rm Ai}''(z)-z {\rm Ai}(z)=0$ and
vanishes as, ${\rm Ai}(z)\approx \frac{1}{2 \sqrt{\pi}z^{1/4}}\, e^{-\frac{2}{3}z^{3/2}}$ as
$z\to \infty$. This solution of the special case $\alpha=0$ of 
the Painlev\'e-II equation
is called the Hastings-McLeod solution~\cite{HM}.   
For $\beta=2$ and $\beta=4$, one has~\cite{TW1,TW2}
\begin{eqnarray}
F_1(x) & =& \left[F_2(x)\right]^{1/2}\, \exp\left[\frac{1}{2}\int_x^{\infty} q(z) dz\right] \label{f1tw} \\
F_4(x)&=& \left[F_2(x)\right]^{1/2}\, \cosh\left[\frac{1}{2}\int_x^{\infty} q(z) dz\right]. \label{f4tw}
\end{eqnarray}  
Note that $F_{\beta}(x)={\rm Prob}(\chi\le x)$ is the cumulative probability of the scaled random variable 
$\chi$ and hence it approaches to $1$ as $x\to \infty$ and vanishes to $0$ as $x\to -\infty$.
The corresponding probability density function (pdf) $F_{\beta}'(x)= dF_{\beta}(x)/dx$ vanishes
as $x\to \pm \infty$ in an asymmetric fashion
\begin{eqnarray}
F_{\beta}'(x) &\sim & \exp\left[-\frac{\beta}{24}\, |x|^3\right] \quad {\rm as}\quad x\to -\infty \label{lefttw} \\
& \sim & \exp\left[- \frac{2\beta}{3}\, x^{3/2}\right] \quad {\rm as}\quad x\to \infty \label{righttw}
\end{eqnarray}
Over the last decade or so, the Tracy-Widom distribution has appeared in a wide variety
of problems ranging from statistical physics and probability theory all the way to
growth models and biological sequence matching problems (for reviews see~\cite{AD,maj-review,forrester,krug}).
These include the longest increasing subsequence or the Ulam problem~\cite{J1,BDJ,AD}, 
a wide variety of (1+1)-dimensional growth 
models~\cite{PS,GTW,J2,MN1,RS1}, 
directed polymer
in random medium~\cite{J3} and the continuum 
Kardar-Parisi-Zhang equation~\cite{SS,CLR,dotsenko,corwin}, Bernoulli 
matching problem between two random sequences~\cite{MN2},
nonintersecting Brownian motions (see e.g. \cite{FMS,RS2} and references 
therein). 
This distribution has also been measured in a variety of recent experiments, e.g., in
the height distribution of fronts generated in paper burning experiment~\cite{pburning}, in
turbulent liquid crystals~\cite{TS} and more recently in coupled fiber laser systems~\cite{Davidson}.

\vspace{0.2cm}

\noindent {\bf Large deviations:} Tracy-Widom distribution describes the probability of {\em typical} fluctuations of $\lambda_{\rm max}$ 
around its mean (on a scale of $N^{-1/6}$), but not the {\em atypical} large fluctuations,
i.e., fluctuations of order
$O(\sqrt{N})$ around the mean value $\sqrt{2 N}$. Questions regarding  
such large/rare fluctuations do arise in various contexts~\cite{deanSM,deanSMlong,vmb}
and have recently been computed~\cite{deanSM,deanSMlong,vmb,vergassolaSM} to dominant order
for large $N$. As a summary, the probability density of $\lambda_{\rm max}$,
$\mathcal{P}\left(\lambda_{\rm max}= t\right)= 
\frac{d}{dt}[\mathbb{P}_N\left(\lambda_{\rm max} \leq t\right)]$,  is given for large $N$ by:
\begin{equation}\hspace{-1.6cm}
 \mathcal{P}\left(\lambda_{\rm max}= t\right)\approx
\left\{\begin{array}{ll}
      \exp\left\{-\beta N^2 \psi_{-}\left(\frac{t}{\sqrt{ N}}\right)+...\right\} 
& \textrm{for $t<\sqrt{2 N}$ and 
$|t-\sqrt{2 N}| \approx O(\sqrt{N})$} \\
& \\
\frac{1}{a_{\beta} N^{-1/6}}\; 
 F_{\beta}'\left(\frac{t-\sqrt{2 N}}{a_{\beta} N^{-1/6}}\right)
& \textrm{for  
$|t-\sqrt{2 N}| \approx O\left(N^{-1/6}\right)$}\\
& \\
     \exp\left\{-\beta N \psi_{+} \left(\frac{t}{\sqrt{ N}}\right)+...\right\} & 
\textrm{for $t>\sqrt{2 N}$ and
$|t-\sqrt{2 N}| \approx O(\sqrt{N})$}
       \end{array}\right.
\label{extremepdf}
\end{equation}
where $F_{\beta}(x)$ is the Tracy-Widom distribution and 
where $\psi_{-}$ and $\psi_{+}$ are respectively the left and right large deviation functions
describing the tails of the distribution of $\lambda_{\rm max}$. 
The rate function  $\psi_{-}(z)$ was explicitly computed in~\cite{deanSM,deanSMlong}, 
while $\psi_{+}(z)$ was computed in~\cite{vergassolaSM}, both by simple physical methods
exploiting the Coulomb gas
analogy. A more complicated, albeit mathematically rigorous, derivation of $\psi_{+}(z)$ in the context of 
spin glass models can be found
in~\cite{benarous}. These rate functions read
\begin{eqnarray}\label{eq:psi_-+}
\psi_{-}(z)&=&\frac{z^2}{3}-\frac{z^4}{108}-\sqrt{z^2+6} \:\; \frac{\left(z^3+15 z\right)}{108}
-\frac{1}{2}\ln\left[\frac{\sqrt{z^2+6}+z}{\sqrt{2}}\right]+\frac{\ln 3}{2}, 
\;\; \textrm{for $z<\sqrt{2}$}\nonumber\\
\psi_{+}(z)&=&\frac{z \sqrt{z^2-2}}{2} +\ln\left[\frac{z-\sqrt{z^2-2}}{\sqrt{2}}\right],\; \;\; \;\;\; \textrm{for $z>\sqrt{2}$.}
\end{eqnarray}
Note that in ref.~\cite{vergassolaSM},
the function $\psi_+(z)$ was expressed in terms of a 
complicated hypergeometric function, which
however can be reduced to a simple algebraic function as 
presented above in Eq. (\ref{eq:psi_-+}).
Note also that while $F_\beta(x)$ depends explicitly on $\beta$, the rate 
functions $\psi_{-}(z)$ and $\psi_+(z)$
are independent of $\beta$.
These rate functions only give the dominant order for large $N$ in the 
exponential.
In other words, the precise meaning of $\approx$ is that for large $N$:  
$\displaystyle{\lim_{N\to \infty} \frac{1}{\beta N^2} \ln \mathcal{P}\left(\lambda_{\rm max}
=z\sqrt{N}\right)= 
-\psi_{-}(z)}$  for $z<\sqrt{2}$
and
$\displaystyle{\lim_{N\to \infty} \frac{1}{\beta N} \ln \mathcal{P}\left(\lambda_{\rm max}=
z \sqrt{N}\right)=
-\psi_+(z)}$ for $z>\sqrt{2 }$.
When $z$ approaches $\sqrt{2}$ (from below or above) it is easy to see that the rate functions 
vanish respectively as
\begin{eqnarray}
\psi_{-}(z) &\to & \frac{1}{6\sqrt{2}} (\sqrt{2}-z)^3+\dots \label{leftv1} \\ 
\psi_+(z) &\to & \frac{2^{7/4}}{3} (z-\sqrt{2})^{3/2}+\dots \label{rightv1} 
\end{eqnarray}  
Note that the physics of the left tail~\cite{deanSM,deanSMlong} is very different
from the physics of the right tail~\cite{vergassolaSM}. In the former case, the
semi-circular charge density of the Coulomb gas is {\em pushed} by the hard wall ($z<\sqrt{2}$)
leading to a reorganization of all the $N$ charges that gives rise to an 
energy difference of $O(N^2)$~\cite{deanSM,deanSMlong}. In contrast, for the right tail 
$z>\sqrt{2}$, the dominant 
fluctuations are caused by {\em pulling} a single charge away (to the right) from the
Wigner sea leading to an energy difference of $O(N)$~\cite{vergassolaSM}.

The different behaviour of the probability distribution for $z<\sqrt{2}$ and $z>\sqrt{2}$ leads
to a `phase transition' strictly in the $N\to \infty$ limit at the critical point $z=\sqrt{2}$
in the following sense. Indeed, if one
scales $\lambda$ by $\sqrt{N}$ and takes the $N\to \infty$ limit, one obtains
\begin{eqnarray}
-\lim_{N\to \infty} \frac{1}{\beta N^2}\, \ln {\mathbb P}_N(\lambda_{\rm max}\le z\sqrt{N}) &= & \psi_{-}(z)\quad 
{\rm for}\quad z< \sqrt{2} \nonumber \\ 
&=& 0 \quad\quad\quad
{\rm for}\quad z> \sqrt{2}
\label{phtrans1}
\end{eqnarray}
Note that since ${\mathbb P}_N(\lambda_{\rm max}\le t)$ can be interpreted as a
partition function of a Coulomb gas (see Eq. (\ref{multint})), its logarithm has
the interpretation of a free energy.
Since $\psi_{-}(z) \sim (\sqrt{2}-z)^{3}$ as $z\to \sqrt{2}$ from below, the $3$-rd 
derivative of the free energy is discontinuous at the critical point $z=\sqrt{2}$.
Hence, this can be interpreted as 
a {\em third order} phase transition.

However, for finite but large $N$, it follows from \eqref{extremepdf} that the behavior to the 
left of $z=\sqrt{2}$ smoothly crosses over to the behaviour to the right as one varies
$z$ through its critical point $z=\sqrt{2}$ and the Tracy-Widom distribution in \eqref{extremepdf}
around the critical point is precisely this crossover function. 
Indeed, if one zooms in close to the mean value $\sqrt{2 N}$ by 
setting $t=\sqrt{2 N}+ x N^{-\frac{1}{6}} /\sqrt{2} $ (for $\beta=1,2$)
in the rate functions $\psi_{-}(t/\sqrt{N})$ and $\psi_{+}(t/\sqrt{N})$ in \eqref{extremepdf},
one expects to recover, by taking large $N$ limit,
respectively the left and the right tail of the Tracy-Widom distribution.
With this scaling, and using \eqref{rightv1}, one finds 
$\psi_{+}(t/\sqrt{N}) \approx \frac{2 x^{3/2}}{3 N}$ and thus
$ \mathcal{P}\left(\lambda_{\rm max}=t\right)\sim \exp\left\{ - \frac{2 \beta}{3} x^{3/2}\right\}$,
which indeed matches the dominant order in the far right tail of the Tracy-Widom distribution 
for $\beta=1,2$ in \eqref{righttw}.
Similarly for the left tail ($x<0$), using \eqref{leftv1}, one finds 
$\psi_{-}(t/\sqrt{N})\approx \frac{|x|^3}{24 N^2}$, thus 
$ \mathcal{P}\left(\lambda_{\rm max}=t\right)\sim \exp\left\{ - \frac{ \beta}{24} |x|^{3}\right\}$
which matches the left tail of the Tracy-Widom distribution in \eqref{lefttw}.

\begin{figure}
 \includegraphics[width=12cm]{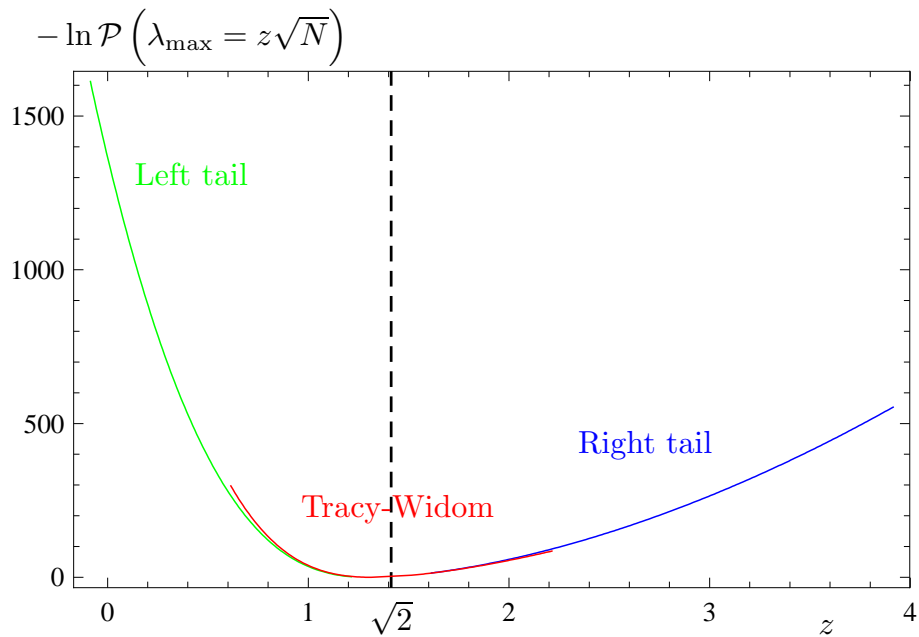}
\caption{Rate function $-\ln\mathcal{P}(\lambda_{\rm max}=z \sqrt{N})$ 
associated to the distribution $\mathcal{P}(\lambda_{\rm max}=t)$ of the maximal eigenvalue of
a random matrix from the GUE for large $N$. Close to the mean value $z=\sqrt{2}$, the distribution
is a Tracy-Widom law (red line), it describes the small typical fluctuations
around the mean value. Atypical large flutuations are described by
the large deviations:
the left large deviation in green ($z<\sqrt{2}$), the right deviation in blue
($z>\sqrt{2}$).}\label{fig:largeDev}
\end{figure}

More recently, higher order corrections for large $N$ have been computed for the 
left tail of the distribution~\cite{borotN}
using methods developed in the context of matrix models.
Note that in ~\cite{borotN} a different notation for $\beta$ was used: $\beta=1/2$ (GOE), $\beta=1$ (GUE)
and $\beta=2$ (GSE). 
To avoid confusion, we present below the results in terms of the standard Dyson index $\beta=1,\, 2,\, 4$.
\begin{equation}
\mathcal{P}\left(\lambda_{\rm max}=t\right) \approx
      \exp\left\{-\Phi_N\left(\frac{t}{\sqrt{N}},\beta\right)\right\}\;\;\;\textrm{for $t<\sqrt{2 N}$ and 
$|t-\sqrt{2 N}| \approx O(\sqrt{N})$}
\end{equation}
where
\begin{equation}\label{eq:phiGaetan}
 \Phi_N(z,\beta)=\beta N^2  \psi_{-}(z)+ N (\beta-2) \Phi_1(z)+\phi_{\beta}\, \ln 
N +\Phi_0(\beta,z)
\end{equation}
with $\psi_{-}(z)$ given in Eq. \eqref{eq:psi_-+} (dominant order). The subleading terms
are given by~\cite{borotN}
\begin{eqnarray}
\Phi_1(z)&=&\frac{z^2}{6}-\frac{z \sqrt{z^2+6}}{12}+
\frac{z}{4 \sqrt{3}} (z^2+6)^{\frac{1}{4}}(\sqrt{z^2+6}-2 z)^{\frac{1}{2}} \nonumber \\
&+&\frac{\ln 18}{4}-\frac{1}{2} \ln\left[2 \sqrt{z^2+6}-z +
\sqrt{3}(z^2+6)^{\frac{1}{4}}(\sqrt{z^2+6}-2 z)^{\frac{1}{2}}\right]
\label{Phi1}
\end{eqnarray}
and
\begin{equation}\label{phibeta}
\phi_{\beta}=-\frac{7}{4}-\frac{1}{12}\left(\frac{\beta}{2}+\frac{2}{\beta}\right),
\end{equation}
and (see Eq. (4-35) in Ref. \cite{borotN})
\begin{eqnarray}
\Phi_0(\beta,z)&=&\left(\frac{1}{12} \beta  +\frac{1}{3 \beta }-\frac{1}{3}\right)
\ln 2+\left(\frac{19}{12 \beta }+\frac{19\beta }{48}+\frac{9}{8}\right)\ln 3 \nonumber \\
&+&\frac{1}{2}\ln \pi 
+\left(\frac{-21}{48}+\frac{11}{24}\left(\frac{1}{\beta }+\frac{\beta }{4}\right) \right) 
\ln \left[ 6+z^2 \right]+\left(\frac{3}{8}-\frac{1}{4 \beta }-\frac{\beta }{16}\right)
 \ln \left[ -2 z+\sqrt{6+z^2} \right]\nonumber \\
&+&\left(\frac{1}{2} -\frac{1}{3\beta }-\frac{\beta }{12}\right)
\ln \left[ z+\sqrt{6+z^2}\right] \nonumber \\
&+&\left(\frac{-4}{3}+\frac{4}{3 \beta } +
\frac{\beta  }{3}\right) 
\ln \left[\sqrt{-2 z+\sqrt{6+z^2}}+\sqrt{3}\left(z^2+6\right)^{1/4} \right]\nonumber \\
&+&\frac{5}{3}\left(1-\frac{1}{\beta }
-\frac{\beta }{4}\right)\ln \left[-z+2 \sqrt{6+z^2}+\sqrt{3}\left(z^2+6\right)^{1/4}
\sqrt{\sqrt{6+z^2}-2 z }\right]\nonumber \\
&-&\ln \left[\left(-18 +z^2\right)z+\left(6+z^2\right)^{3/2}\right]-
\frac{\ln \beta }{2}-\kappa_{\beta} 
\label{Phi0}
\end{eqnarray}
where $\kappa_{\beta}$ is a complicated function of $\beta$.
For $\beta/2$ integer, it reduces to~\cite{borotN}
\begin{equation}
\kappa_{\beta}=\left(\frac{\beta /2+1}{4}\right)\ln\left(2 \pi \right)+
\frac{2 \zeta'\left(-1\right)}{\beta } 
-\frac{\ln\left(\beta /2\right)}{6 \beta }-
\sum_{m=1}^{\frac{\beta}{2}-1}\frac{2 m}{\beta}\ln\Gamma\left(2 \frac{m}{\beta }\right).
\label{kappa2}
\end{equation}
For instance, for the GUE ($\beta=2$), we find $\kappa_{2}=\frac{\ln\left(2 \pi\right)}{2}+\zeta'(-1)$.
For $\beta=1,2$ and $4$, the expression in Eq. \eqref{eq:phiGaetan} matches the left asymptotics of the Tracy-Widom
distribution, i.e. the asymptotic behaviour of $F_{\beta}'(x)$ for $x\rightarrow -\infty$, 
see ref. \cite{asymptTWleft}.
However, for the right tail of the distribution of $\lambda_{\rm max}$, the corrections 
to dominant order for large $N$ 
are, to our knowledge, not known until now. 
In fact, one of the results of this paper is to compute these right tail corrections for the GUE ($\beta=2$).
Both left and right large deviations are plotted in Fig. \ref{fig:largeDev} for the GUE.
The left tail is described by $\Phi_N(z,2)$ in Eq. \eqref{eq:phiGaetan}, the right tail is described
by our result given in Eq. \eqref{eq:rightTailIntro}.

Another result of this paper concerns a simpler and pedestrian derivation of the Tracy-Widom
distribution for the GUE case.
The original derivation of the Tracy-Widom law for the distribution of  
typical fluctuations of $\lambda_{\rm max}$~\cite{TW1,TW2} is somewhat complex
as it requires a rather sophisticated and nontrivial asymptotic analysis
of the Fredholm determinant involving Airy Kernel~\cite{TW1,TW2}. Since this distribution
appears in so many different contexts, it is  quite natural to ask if there is any
other simpler (more elementary) derivation of the Tracy-Widom distribution.
In this paper, we provide such a derivation for the GUE case. Our method
is based on a suitable modification of a technique of orthogonal polynomials
developed by Gross and Matytsin~\cite{gross} in the context of two-dimensional
Yang-Mills theory. In fact, the partition function of the continuum two-dimensional 
pure Yang-Mills
theory on a sphere (with gauge group ${\rm U}(N)$) can be written (up to a prefactor)
as a discrete multiple 
sum over integers~\cite{Migdal,Rusakov}
\begin{equation}
Z(A,N)= \sum_{n_1,n_2,\dots, n_N=-\infty}^{\infty}\;  \prod_{1\le i<j\le N}(n_i-n_j)^2\, 
e^{-(A/2N)\sum_{j=1}^N n_j^2}
\label{ympf}
\end{equation}
where $A$ is the area of the sphere. 
In the $N\to \infty$ limit, the free energy $\ln Z$, as a function of $A$, undergoes
a 3rd order phase transition known as the Douglas-Kazakov transition~\cite{DK} at the
critical value $A_c=\pi^2$. For $A>A_c$, the system is in the `strong' coupling phase
while for $A<A_c$, it is in the `weak' coupling phase. For finite but large $N$, there
is a crossover between the two phases as one passes through the vicinity of the critical 
point. In the so called double scaling limit (where $A\to A_c$, $N\to \infty$ but keeping
the product $(A-A_c)N^{2/3}$ fixed), the singular part of the free energy satisfies
a Painlev\'e II equation~\cite{gross}.
Gross and Matytsin (see also ~\cite{cresc}) used a method based on orthogonal
polynomials to analyse the partition sum in the double scaling limit, as well as
in the weak coupling regime ($A<A_c$) where they were able to compute non-perturbative (in $1/N$ expansion)
corrections to the free energy. 
Actually, a similar $3$-rd order phase transition from a weak to 
strong coupling phase
in the $N\to \infty$ limit was originally noticed in the lattice formulation (with Wilson action) 
of the two dimensional
${\rm U}(N)$ gauge theory~\cite{WG,Wadia,J1} and in the vicinity of the transition point the
singular part of the free energy was shown to 
satisfy a Painlev\'e II equation~\cite{periwal}.  
Note that similar calculations involving the asymptotic analysis
of partition functions using orthogonal polynomials were
used extensively in the early 90's to study the double scaling 
limit of the so called one-matrix model (for a recent review and 
developments, see e.g. Ref. ~\cite{Marino}).

In our case, for the distribution of $\lambda_{\rm max}$, we need to analyse 
the asymptotic large $N$ behaviour of  
a multiple {\em indefinite} integral in Eq. (\ref{multint}), as opposed to the discrete
sum in Eq. (\ref{ympf}). However, we show that one can suitably modify the orthogonal
polynomial method of Gross and Matytsin to analyse the multiple integral in Eq. (\ref{multint})
in the limit of large $N$. In fact, we find a similar third order phase transition (in the $N\to \infty$
limit) in the largest eigenvalue distribution $\mathbb{P}_N\left(\lambda_{\rm max} \leq t\right)$ as
a function of $t$ at the critical point $t_c= \sqrt{2N}$.  
The regime of left large deviation
of $\mathbb{P}_N\left(\lambda_{\rm max} \leq t\right)$ ($t<t_c$) is similar
to the `strong' coupling regime $(A>A_c)$ of the Yang-Mills theory, while the
right large deviation tail of $\mathbb{P}_N\left(\lambda_{\rm max} \leq t\right)$ ($t>t_c$)
is similar to the `weak' coupling regime $(A<A_c)$ of the Yang-Mills theory. 
For finite but large $N$,
the crossover function across the critical point that connects the 
left and right large deviation tails
is precisely the Tracy-Widom distribution. Thus the Tracy-Widom distribution corresponds
precisely to the double scaling limit of the Yang-Mills theory and one finds the
same Painlev\'e II equation. A similar 3rd order phase transition was also found
recently in a model of non-intersecting Brownian motions by establishing an exact
correspondence between the reunion probability
in the Brownian motion model and the partition function in the 
$2$-d Yang-Mills theory on a sphere~\cite{FMS,SMCR}.  

The advantage of this orthogonal polynomial method to analyse the
maximum eigenvalue distribution is twofold: (i) one gets
the Tracy-Widom distribution in a simple elementary way (basically one carries out
a scaling analysis of a pair of nonlinear recursion relations near the critical point 
and shows that the scaling function
satisfies a Painlev\'e II differential equation) and (ii) as an added bonus, we also obtain precise 
subleading corrections to the leading right large deviation tail $(t>\sqrt{2N})$. The subleading corrections,
in the Yang-Mills language, correspond to the non-perturbative corrections in the
weak coupling regime as derived by Gross and Matytsin~\cite{gross}. More precisely
we show that       
\begin{equation}\label{eq:rightTailIntro}
\mathcal{P}\left(\lambda_{\rm max} =t \right)
\approx \frac{\sqrt{N}}{ 2 \pi \sqrt{2 }\, ( t^2-2 N)} \: e^{-2 N \psi_+\left(\frac{t}{\sqrt{N}}\right)}\;\;
{\rm for}\;\; t>\sqrt{2 N}\;\;,\;\;
\left|t-\sqrt{2 N}\right| \approx O(\sqrt{N})\;\:\;\;\;
\end{equation}
where $\psi_{+}(z)$ is given in Eq. (\ref{eq:psi_-+}). Note that only the leading behaviour
$\exp\left[-2N \psi_{+}(z)\right]$ was known before~\cite{vergassolaSM}, but the subleading corrections are,
to our knowledge, new results. 
We also verify that our expression matches the precise right asymptotics of the Tracy-Widom distribution.
Figure \ref{fig:largeDev} shows the distribution of  $\lambda_{\rm max}$
for the GUE: close to the mean value it is described by the Tracy-Widom distribution,
whereas the tails are described by the large deviations. The right tail (right large deviation)
is given by our result in Eq. \eqref{eq:rightTailIntro}.
Together with the subleading terms in the left tail in Eq. (\ref{eq:phiGaetan}),
our new result in Eq. (\ref{eq:rightTailIntro}) then provides a rather complete picture of the 
tail behaviors of the distribution
of $\lambda_{\rm max}$ on both sides of the mean $\sqrt{2N}$.

The rest of the paper is organized as follows.
In Section \ref{sec:Intro}, we start with some general notations and scaling remarks for the GUE.
In Section \ref{sec:Orthpol}, we explain the method of orthogonal polynomials on a semi-infinite interval
and derive some basic recursion relations.
In Section \ref{sec:LargeDev}, we compute the right tail of the distribution of $\lambda_{\rm max}$
(dominant order and corrections for the GUE): it describes atypical large fluctuations of $\lambda_{\rm max}$ 
to the 
right of its mean value.
In Section \ref{sec:TW}, using results of the previous sections and basic scaling remarks,
we derive the Tracy-Widom law (with $\beta=2$ for the GUE) that describes small typical fluctuations 
close to the mean value.

\section{Notations and scaling}
\label{sec:Intro}

In the rest of the paper we focus only on Gaussian random  matrices $X$ drawn from the GUE ($\beta=2$).
They are Hermitian random matrices $X^{\dagger}=X$
such that
$\mathcal{P}(X)\propto e^{-\alpha \, {\rm Tr}\left(X^2 \right)}$
where we have introduced an additional parameter $\alpha>0$ for the purpose
of certain mathematical manipulations that will be clear later.
Setting $\alpha=1$ at the end of the calculation, we will recover the usual GUE.
With the additional parameter $\alpha$, the joint distribution of the eigenvalues of $X$ 
is given by (see Eq. \eqref{eq:jpdfEV}):
\begin{equation}\label{eq:JointPdfEv}
\mathcal{P}\left(\lambda_1,...,\lambda_N\right)=
B_N(\alpha)\; e^{- \alpha\, \sum_{i=1}^N \lambda_i^2}\: \prod_{j<k} \left(\lambda_j-\lambda_k\right)^2
\end{equation}
where $B_N(\alpha)=(2 \alpha)^{\frac{N^2}{2}}\, (2 \pi)^{-\frac{N}{2}}/\left[\prod_{n=1}^N n! \right]$ is 
the normalization constant.
The Vandermonde determinant appears with a power $2$ as we consider the GUE ($\beta=2$).
This determinant indeed comes from a Jacobian for the change of variables from the entries of the matrix to its eigenvalues.
The power is related to the number of real degrees of freedom of an element of the matrix, which is $2$ for 
complex entries, i.e., for GUE
(it is $1$ for real entries GOE and $4$ for quaternion entries GSE). As we will see later, 
this power $2$ is crucial
for the method of orthogonal polynomials to work. 
The technique of Gross and Matytsin~\cite{gross} that we adapt here also works only for the 
GUE $\beta=2$ case.
\\

Given the joint distribution of eigenvalues in Eq. (\ref{eq:JointPdfEv}), it is easy to
write down the
cumulative distribution of the maximal eigenvalue $\lambda_{\rm max}$ 
\begin{equation}\label{eq:CdfLmax}
\mathbb{P}_N\left(\lambda_{\rm max} \leq y\right) ={\rm Prob}\left[\lambda_1\le y, \lambda_2\le y,\dots, \lambda_N\le y\right]=\frac{Z_N \left(y,\alpha\right)}{Z_N(\infty,\alpha)}
\end{equation}
where the partition function $Z_N$ is given by the multiple indefinite integral
\begin{equation}\label{eq:Zn}
Z_N(y,\alpha)= \frac{1}{N!} \prod_{i=1}^N \int_{-\infty}^y d\lambda_i\: \:\prod_{j<k} \left(\lambda_j-\lambda_k\right)^2\:
e^{- \alpha\, \sum_{i=1}^N \lambda_i^2}
\end{equation}
The normalization $Z_N(\infty,\alpha)$ is actually related to $B_N(\alpha)$
in Eq. \eqref{eq:JointPdfEv} as $B_N(\alpha)=1/\left(N!\, Z_N(\infty,\alpha)\right)$.
Note that, by the trivial rescaling $\sqrt{\alpha} \lambda_i\to \lambda_i$ in \eqref{eq:Zn},
it follows from \eqref{eq:CdfLmax} that
\begin{equation}
\label{rescaled1}
\mathbb{P}_N\left(\lambda_{\rm max} \leq y\right) =\frac{Z_N(y,\alpha)}{Z_N(\infty,\alpha)}
=\frac{Z_N(y\sqrt{\alpha},1)}{Z_N(\infty,1)}.
\end{equation}
Thus $y$ and $\alpha$ always appear in a single scaling combination $y\sqrt{\alpha}$.

We will henceforth focus on the large $N$ limit. For fixed $\alpha$, one can easily
figure out from the joint pdf in Eq. (\ref{eq:JointPdfEv}) how a typical eigenvalue 
${\rm \lambda}_{\rm typ}$ scales with $N$ for large $N$.
Let us rewrite the joint distribution
of eigenvalues in Eq. \eqref{eq:JointPdfEv} as 
\begin{equation}
\mathcal{P}(\lambda_1,...\lambda_N) \propto \exp\left\{ -\alpha \sum_i \lambda_i^2 +2 \sum_{j<k } 
\ln|\lambda_j-\lambda_k| \right\}
\label{coulombgas}
\end{equation}
which can then be interpreted as a Boltzmann weight $\propto \exp\left[-E_{\rm eff}\right]$, 
with effective energy $E_{\rm eff}=\alpha \sum_i \lambda_i^2 -2 \sum_{j<k } \ln|\lambda_j-\lambda_k|$.
The eigenvalues can thus be seen as the positions of $N$ charges of a 2D Coulomb gas (but restricted
to be on the real line) which repel each other 
via a logarithmic
Coulomb potential (coming from the Vandermonde determinant in Eq. \eqref{eq:JointPdfEv})~\cite{forrester}.
In addition, the charges are subjected to an external confining parabolic potential.
For large $N$, the first term of the energy is of order $N \lambda_{\rm typ}^2$, whereas the 
second is of order $N^2$ because of the double sum.
Balancing the two terms $N \lambda_{\rm typ}^2 \sim N^2$ gives
the scaling of a typical eigenvalue for large $N$: $\lambda_{\rm typ} \sim \sqrt{N}$.
For large $N$, the eigenvalues are close to each other and they can be described 
by a continuous charge density (normalized to unity)
$\rho_N(\lambda)=\frac{1}{N} \sum_i \delta\left(\lambda -\lambda_i \right)$.
The average density of states for large $N$ 
can be obtained by evaluating the full partition function $Z_N(\infty, \alpha)$ (the denominator
in Eq. (\ref{eq:CdfLmax})) via a saddle point method. The saddle point density
is the density that minimizes the effective energy (see  the book of Mehta ~\cite{mehta})
$E_{\rm eff}=\alpha N \int d\lambda \rho_N(\lambda) \lambda^2- N^2 \int d\lambda \int d\lambda' \rho_N(\lambda) \rho_N(\lambda') \ln|\lambda-\lambda'|$
(in its continuous version). This gives the well-known semi-circle law 
(which is exactly the same as in Eq. \eqref{eq:averDens} for $\alpha=1$):
\begin{equation}
\rho_N(\lambda) = \frac{1}{\sqrt{N}} \, \rho\left(\frac{\lambda}{\sqrt{N}}\right)
 \;\; {\rm with }\;\; \rho(x)= \frac{\alpha}{ \pi } \sqrt{\frac{2 }{\alpha}-x^2}
\end{equation}
The density is plotted in figure \ref{fig:densTW}.
\\

The average value of $\lambda_{\rm max}$ is  again given for large $N$ by
the upper bound of the density support (see Fig. \ref{fig:densTW}):
\begin{equation}
 \langle  \lambda_{\rm max} \rangle\approx \sqrt{\frac{2 N}{\alpha}}\;\;\; \textrm{for large $N$}
\end{equation}
For $\alpha=1$, this evidently reduces to the usual expression for $\langle \lambda_{\rm max}\rangle$.
The typical scaling for large $N$
is  thus $\lambda_{\rm max} \sim \sqrt{N}$. Hence, we will use $\lambda_{\rm max}=z\, \sqrt{N}$, 
where $z$ is of order one.

\section{Orthogonal polynomials}
\label{sec:Orthpol}

In this section, we introduce the method of orthogonal polynomials
to evaluate the partition function in Eq. (\ref{eq:Zn}).
As mentioned in the introduction, to evaluate this multiple indefinite integral
we will adapt the method developed 
by Gross and Matytsin~\cite{gross} to enumerate the partition sum \eqref{ympf} in the
two-dimensional Yang-Mills theory.
Orthogonal polynomials
are very useful to handle the square Vandermonde determinant in
the distribution of the eigenvalues in Eq. \eqref{eq:JointPdfEv}.
A Vandermonde determinant can indeed be written as
$\prod_{i<j }(\lambda_j -\lambda_i) = \det\left(\lambda_i^{j-1}\right)_{i,j}=
\det\left(p_{j-1}(\lambda_i)\right)_{i,j} $ where  $p_j(\lambda)=\lambda^j+...$
is any polynomial of degree $j$ with leading coefficient one.
The idea is to choose well these polynomials $p_j$ in order to simplify the computation
of the integral.

We define an operation on pairs of polynomials as follows:
\begin{equation} \label{eq:ScalProd}
\langle f ,g \rangle=\int_{-\infty}^y d\lambda\, e^{-\alpha \lambda^2}\; f\left(\lambda\right) g\left(\lambda\right)
\end{equation}
We consider a family  $\left\{p_n\right\}_{n\geq 0}$ of orthogonal polynomials
with respect to the operation defined above, i.e. with weight $e^{-\alpha \lambda^2}$
on the interval $]-\infty,y]$. Without any loss of generalization, we define the 
polynomial $p_n(\lambda)$ of degree $n$
such that the coefficient of $\lambda^n$ term is always fixed to be $1$, i.e.,
$p_n(\lambda)=\lambda^n+...$. These polynomials satisfy the orthogonality
property: $\langle p_n,p_m\rangle = 0$ for all $n\neq m$.
We also write $h_n =\langle p_n,p_n\rangle $. Thus,
\begin{equation}
 \langle p_n,p_m\rangle=\delta_{n,m}\: h_n\;\;\textrm{for all}\;\; n\geq 0
\end{equation}
Note that $p_n(\lambda)$ and $h_n$ are implicitly functions of $\alpha$ and $y$, i.e.
$p_n(\lambda)=p_n\left(\lambda|y,\alpha\right)$ and $h_n=h_n(y,\alpha)$.
In particular, we can easily compute by hand the first few $p_n(\lambda)$'s for fixed $\alpha >0$
and $y$, but the expressions become rather complex as $n$ increases and it is hard to
find a closed form expression for $p_n(\lambda)$ for every $n$
(except for the limiting case $y\rightarrow \infty$): 
\begin{eqnarray*}
p_0(\lambda)&=&1  \\
p_1(\lambda)&=&\lambda+\frac{e^{-\alpha y^2}}{\sqrt{\pi \alpha } \left[1+{\rm erf}(y \sqrt{\alpha})\right]}\\
p_2(\lambda)&=&\lambda^2+\frac{-2 y \sqrt{\alpha }-e^{y^2 \alpha } \sqrt{\pi } 
\left(1+2 y^2 \alpha \right) \left(1+\text{erf}\left[y \sqrt{\alpha }\right]\right)
}{\sqrt{\alpha }
 \left(2+2 e^{y^2 \alpha } \sqrt{\pi } y \sqrt{\alpha } \left(1+\text{erf}\left[y \sqrt{\alpha }\right]\right)
-e^{2 y^2 \alpha } \pi  \left(1+\text{erf}\left[y \sqrt{\alpha }\right]\right)^2\right)} \; \lambda+\\
&& \;\;\;\;\;\;\;\;\;+ \frac{-4-4 e^{y^2 \alpha } \sqrt{\pi } y \sqrt{\alpha } \left(1+\text{erf}\left[y \sqrt{\alpha }\right]\right)
+e^{2 y^2 \alpha } \pi  \left(1+\text{erf}\left[y \sqrt{\alpha }\right]\right)^2
}{4 \alpha +4 e^{y^2 \alpha } \sqrt{\pi } y \alpha ^{3/2} \left(1+\text{erf}
\left[y \sqrt{\alpha }\right]\right)-2 e^{2 y^2 \alpha } 
\pi  \alpha  \left(1+\text{erf}\left[y \sqrt{\alpha }\right]\right)^2}
\end{eqnarray*}
Thus we get for instance
$h_0=\langle p_0|p_0\rangle=\frac{\sqrt{\pi}}{2\sqrt{\alpha}}\left[1+{\rm erf}(y \sqrt{\alpha})\right]$
\\
and $h_1=\langle p_1 |p_1 \rangle=\frac{-2 y \sqrt{\alpha} \: e^{-\alpha y^2}+\sqrt{\pi}\left[1+{\rm erf}(y \sqrt{\alpha})\right]
-\frac{2 e^{-2 \alpha y^2}}{\sqrt{\pi}\left[1+{\rm erf}(y \sqrt{\alpha})\right]}
}{4 \alpha^{3/2}}$, etc.
In the limit $y\rightarrow \infty$, we recover the Hermite polynomials:
$p_0=1$, $p_1=\lambda$, $p_2=\lambda^2-\frac{1}{2 \alpha}$
and $h_0=\frac{\sqrt{\pi}}{\sqrt{\alpha}}$, $h_1=\frac{\sqrt{\pi}}{2 \alpha^{3/2}}$,
$h_2=\frac{\sqrt{\pi}}{2 \alpha^{5/2}}$.

\subsection{Partition function}

The partition function $Z_N(y,\alpha)$ in Eq. \eqref{eq:Zn}
can be written as a function of the $h_n$'s.
By combination of rows, the Vandermonde determinant in the joint distribution of the eigenvalues
can indeed be written
\begin{equation*}
\label{det1}
 \prod_{j<k} (\lambda_k-\lambda_j)=\det\left(\lambda_i^{j-1} \right)_{i,j}=
\det\left(p_{j-1}(\lambda_i) \right)_{i,j}
\end{equation*}
 Then, the partition function can be expressed as
\begin{eqnarray*}
Z_N(y,\alpha)&=&\frac{1}{N!} \prod_{i=1}^N \int_{-\infty}^y d\lambda_i\: \:\prod_{j<k} \left(\lambda_j-\lambda_k\right)^2\:
e^{- \alpha\, \sum_{i=1}^N \lambda_i^2}\\
&=&\frac{1}{N!} \prod_{i=1}^N \int_{-\infty}^y d\lambda_i\: \:
\det\left(p_{j-1}(\lambda_i) \right)_{i,j}\:\det\left(p_{l-1}(\lambda_k) \right)_{k,l}
\:
e^{- \alpha\, \sum_{i=1}^N \lambda_i^2}\\
&=&
\det\left[\int_{-\infty}^y d\lambda \:e^{-\alpha \lambda^2}\: p_{i-1}(\lambda) p_{j-1}(\lambda)\right]_{i,j}\\
&=&\det\left(\langle p_{i-1}|p_{j-1}\rangle\right)_{i,j}
=\det\left(\delta_{i,j} h_{i-1}\right)_{i,j}=\prod_{i=0}^{N-1} h_i
\end{eqnarray*}
where in going from the second to the third line we have used the Cauchy-Binet formula~\cite{mehta}.
Note that this step works only for $\beta=2$.
Therefore the partition function reduces to:
\begin{equation}\label{eq:Znh}
 Z_N(y,\alpha)=\prod_{n=0}^{N-1} h_n(y,\alpha)
\end{equation}
Thus the integral $Z_N(y,\alpha)$ is now expressed as a product of the coefficients $h_n$.
The goal of next subsection is to find recursion relations for the $h_n$'s in order to compute them
and subsequently analyse their product $Z_N$ in Eq. (\ref{eq:Znh})
in the large $N$ limit.

\subsection{Recursion relations}

In general, for orthogonal polynomials
(with any reasonable weight function), one can write a recursion relation of the form:
\begin{equation}\label{eq:recPol}
 \lambda\, p_n(\lambda)=p_{n+1}(\lambda)+S_n\, p_n(\lambda)+R_n \, p_{n-1}(\lambda)
\end{equation}
where $S_n$ and $R_n$ are real coefficients.
This relation comes from the fact that $p_n=\lambda^n+...$ and that
$\langle p_n|q \rangle=0$ for any polynomial $q(\lambda)$
of degree strictly smaller than $n$.
The coefficients $S_n$ and $R_n$ are functions  of $\alpha$ and $y$, i.e.
$S_n=S_n(y,\alpha)$ and $R_n=R_n(y,\alpha)$.
\\

Let us first demonstrate that the coefficients $R_n$ and $S_n$ can be expressed in terms of $h_n$'s.
From Eq. \eqref{eq:recPol}, we  get:
$\langle p_{n-1}|\lambda p_n\rangle=R_n \, \langle p_{n-1}| p_{n-1}\rangle=R_n\, h_{n-1}$.
On the other hand, we have 
$\langle p_{n-1}|\lambda p_n\rangle=\langle \lambda p_{n-1}| p_n\rangle
=\langle p_{n}+S_{n-1}\, p_{n-1}+R_{n-1} \, p_{n-2}| p_n\rangle=\langle p_{n}| p_{n}\rangle=h_n$.
Therefore $R_n h_{n-1}=h_n$, thus $R_n=h_n/h_{n-1}$.
\\

From Eq. \eqref{eq:recPol} again, we also get:
$\langle p_{n}|\lambda p_n\rangle=S_n\, \langle p_n|p_n\rangle=S_n\, h_n$.
By definition we have $\langle p_{n}|\lambda p_n\rangle=\int_{-\infty}^y d\lambda\, e^{-\alpha \lambda^2}\; 
\lambda\, p_n^2\left(\lambda\right)$.
After integrating by part we find:
$\langle p_{n}|\lambda p_n\rangle=-\frac{1}{2 \alpha} e^{-\alpha y^2} p_n^2(y)=-\frac{1}{2 \alpha}
\frac{\partial  h_n(y,\alpha)}{\partial y}$. The last equality follows from
the definition of $h_n$.
As
$h_n=\int_{-\infty}^y d\lambda\, e^{-\alpha \lambda^2}\;  p_n^2\left(\lambda\right)$,
we have indeed $\frac{\partial  h_n(y,\alpha)}{\partial y}=e^{-\alpha y^2} p_n^2\left(y|y,\alpha\right)
+2 \langle p_n|\frac{\partial  p_n}{\partial y}\rangle $. However,  
$\langle p_n|\frac{\partial  p_n}{\partial y}\rangle=0$ as $\frac{\partial  p_n}{\partial y}$
is a polynomial of degree strictly smaller than $n$ (as $p_n\left(\lambda|y,\alpha\right)=\lambda^n+...$).
Therefore $S_n=-\frac{1}{2 \alpha}\, \frac{\partial \ln h_n}{\partial y}$.
\\

Combining these relations, $R_n$ and $S_n$ are then given by:
\begin{equation}\label{eq:RS}
 R_n(y,\alpha)=\frac{h_n(y,\alpha)}{h_{n-1}(y,\alpha)}\;\;
{\rm and}\;\; S_n(y,\alpha)=-\frac{1}{2 \alpha}\, \frac{\partial \ln h_n(y,\alpha)}{\partial y}
\end{equation}
Iterating the recursion relation $h_n=R_n h_{n-1}$ starting from $n=1$, we can write $h_n$ in terms of 
$R_k$'s:
\begin{equation}\label{eq:hn}
 h_n=\left(\prod_{k=1}^n R_k\right) h_0
\end{equation}
Substituting this result in Eq. (\ref{eq:Znh}), we see that the partition function $Z_N$ can be entirely 
expressed in terms of a product over the $R_n$'s. Thus, if we can determine $R_n$'s, we can evaluate
the partition function explicitly. 
\\

Thus the next task is to determine the $R_n$'s. To do this, we will first derive
a set of coupled recursion relations for $R_n$'s and $S_n$'s.
We have $\frac{\partial  h_n}{\partial \alpha}=
\frac{\partial  \langle p_n|p_n \rangle}{\partial \alpha}=
-\int_{-\infty}^y d\lambda\, e^{-\alpha \lambda^2}\; \lambda^2\, p_n^2\left(\lambda\right)
=-\langle \lambda p_n|\lambda p_n\rangle$ where we have used the fact that
$\langle p_n|\frac{\partial p_n}{\partial \alpha}\rangle=0$ (which follows
from the observation that $\frac{\partial p_n}{\partial \alpha}$ is a polynomial
of degree strictly less than $n$ and hence orthogonality dictates that it is identically 
zero).
On the other hand, using Eq. \eqref{eq:recPol} we find:
$\langle \lambda p_n|\lambda p_n\rangle=\langle p_{n+1}+S_n p_n+R_n p_{n-1}|p_{n+1}+S_n p_n+R_n p_{n-1}
\rangle=h_{n+1}+S_n^2 h_n+R_n^2 h_{n-1}=h_n\left(R_{n+1}+S_n^2+R_n\right)$.
Therefore,
\begin{equation}\label{eq:dh}
 -\frac{\partial \ln h_n}{\partial \alpha}=R_n+R_{n+1}+S_n^2
\end{equation}
We can eliminate $h_n$
from the relations \eqref{eq:RS} and \eqref{eq:dh} and get a pair of coupled nonlinear recursion
relations
for $R_n$ and $S_n$.
Using Eq. \eqref{eq:dh} for $n$ and $n-1$, and as $R_n=h_n/h_{n-1}$, we find
$-\frac{\partial \ln R_n}{\partial \alpha}=R_{n+1}-R_{n-1}+S_n^2-S_{n-1}^2$.
Using Eq. \eqref{eq:RS}, we also find $\frac{\partial \ln R_n}{\partial y}=2 \alpha \left(S_{n-1}-S_n \right)$.
Finally, we then get our desired recursion relations:
\begin{eqnarray}
 R_{n+1}&=&-\frac{\partial \ln R_n}{\partial \alpha} +R_{n-1}-S_n^2+S_{n-1}^2 
 \label{eq:daRn}\\
S_n&=&S_{n-1}-\frac{1}{2 \alpha} \frac{\partial \ln R_n}{\partial y} \label{eq:dyRn}
\end{eqnarray}
It is easy to show by induction that
the two relations \eqref{eq:daRn} and \eqref{eq:dyRn}
together with the  initial conditions $R_0$, $R_1$ and $S_0$
uniquely determine $R_n$ and $S_n$.
The additional initial condition $h_0$
is enough to determine $h_n$ as given in Eq. \eqref{eq:hn}.
\\

By definition, $p_0$ is a polynomial of degree $0$
with dominant coefficient $1$. Thus $p_0(\lambda|y,\alpha)=1$.
Therefore $h_0(y,\alpha)=\langle p_0|p_0 \rangle=\int_{-\infty}^y d\lambda \: e^{-\alpha \lambda^2}
=\frac{1}{2}\sqrt{\frac{\pi}{\alpha}}\left(1+{\rm erf}(y \sqrt{\alpha})\right)$.
We also have $R_0(y,\alpha)=0$ as the recursion relation in Eq. \eqref{eq:recPol}
must reduce for $n=0$ to
$ \lambda\, p_0(\lambda)=p_{1}(\lambda)+S_0\, p_0(\lambda)$,
i.e., $p_{1}(\lambda)=\lambda-S_0$.
Moreover we get from Eq. \eqref{eq:RS} 
$S_0=-\frac{1}{2 \alpha}\, \frac{\partial \ln h_0}{\partial y}=-e^{-\alpha y^2}/(2 \alpha h_0)
=-e^{-\alpha y^2}/\left[\sqrt{\pi \alpha} \left(1+{\rm erf}(y \sqrt{\alpha})\right) \right]$.
We now have $R_0$, $S_0$ and $h_0$, we can thus determine $R_1$ from Eq. \eqref{eq:dh} for $n=0$:
$R_1=-\frac{\partial \ln h_0}{\partial \alpha}-R_0-S_0^2=y S_0 +1/(2 \alpha)-S_0^2$.
Thus, the initial conditions can be summarized as:
\begin{eqnarray}\label{eq:CI}
 R_0(y,\alpha)&=&0 \;,\;\;\;\;\;\; p_0(\lambda|y,\alpha)=1\;,\;\;\;\;\;\;\; \nonumber \\
h_0(y,\alpha)&=&\int_{-\infty}^y d\lambda \: e^{-\alpha \lambda^2}=\frac{1}{2}\sqrt{\frac{\pi}{\alpha}}\left(1+{\rm erf}(y \sqrt{\alpha})\right)\nonumber\\
 S_0(y,\alpha)&=&-\frac{1}{2 \alpha}\, \frac{\partial \ln h_0}{\partial y}=-\frac{e^{-\alpha y^2}}{2 \alpha h_0}
=-\frac{e^{-\alpha y^2}}{\sqrt{\pi \alpha} \left(1+{\rm erf}(y \sqrt{\alpha})\right) } \nonumber\\
R_1(y,\alpha)&=& y S_0 +\frac{1}{2 \alpha}-S_0^2
\end{eqnarray}

\subsection{Normalization: limit $y\rightarrow \infty$}

As $y\rightarrow \infty$, we can explicitly compute the functions
$S_n$, $R_n$ and $h_n$. As mentioned above, in the limit $y\rightarrow \infty$ the polynomials
$p_n$ are indeed simply the Hermite polynomials. Hence everything can then be computed explicitly in this case.
We have  $h_0(\infty,\alpha)=\int_{-\infty}^{\infty} d\lambda \: e^{-\alpha \lambda^2}
=\sqrt{\frac{\pi}{\alpha}}$ and $S_0(\infty,\alpha)=0$.
Then, by recurrence it is easy to show that:
\begin{equation}\label{eq:SnRnNorm}
 S_n(\infty,\alpha)=0\;\;\;\;\;\;\;\; R_n(\infty,\alpha)=\frac{n}{2 \alpha}
\end{equation}
Finally, using Eq. \eqref{eq:hn} we get
\begin{equation}
 h_n=\sqrt{\frac{\pi}{\alpha}}\, \frac{n!}{(2 \alpha)^n}
\end{equation}
and thus (see Eq. \eqref{eq:Znh})
\begin{equation}
Z_N(\infty,\alpha)=(2 \pi)^{\frac{N}{2}} (2 \alpha)^{-\frac{N^2}{2}}\prod_{n=1}^N \Gamma(n)
\end{equation}
which could also have been computed directly using Selberg's integral.
We recover the normalization $B_N=1/\left(N!\, Z_N\right)$ in Eq. \eqref{eq:JointPdfEv}.

\section{Right tail of the distribution of $\lambda_{\rm max}$:
large deviation function}
\label{sec:LargeDev}

In the previous section we derived a pair of coupled recursion equations  
\eqref{eq:daRn} and \eqref{eq:dyRn}
with initial conditions given in Eq. \eqref{eq:CI} that determines uniquely
$R_n$, $S_n$ and thus $h_n$ and subsequently $Z_N$ via \eqref{eq:Znh}.
However, these equations are hard to solve explicitly for general $n$ and $y$ 
(apart from the case $y=+\infty$ as shown in the previous subsection).
In this section, we derive an approximate solution for $Z_N$ and hence
for the cdf \eqref{eq:CdfLmax}, in the large $N$ limit
where $N$ is the number of eigenvalues of the matrix $X$ and we will see that
this solution for cdf is valid only for 
$\lambda_{\rm max}> \langle \lambda_{\rm max}\rangle=
\sqrt{\frac{2N}{\alpha}}$, i.e., it only describes the right tail of the 
probability distribution.

We have seen in Section 2 that for large $N$ and fixed $\alpha$, the maximal eigenvalue typically scales as
$\lambda_{\rm max}\sim \sqrt{N}$. We are going to work on this scale, hence in the
definition of the maximal eigenvalue cdf in \eqref{eq:CdfLmax} and \eqref{eq:Zn}, we will set $y=z\sqrt{N}$ 
with $z\approx O(1)$. We will then work in the limit of large $N$ with $z$ fixed.
With this scaling, the operation $\langle f,g\rangle$
defined in Eq. \eqref{eq:ScalProd} for polynomials
depends on $N$ (since the upper limit of integration in \eqref{eq:ScalProd} is now $z\sqrt{N}$). 
The coefficients $R_n$, $S_n$ and $h_n$, for a given $n$, are also now implicitly functions of
$N$. We can make an expansion of these parameters for large $N$ and fixed $n$.
The dominant order will be given  by the $y=+\infty$ case (as in previous subsection)
as $y=z \sqrt{N}\to \infty$ as $N\to \infty$ for fixed $z$. In this section, we want to 
determine the first correction to this dominant term.

However the partition function $Z_N$ and thus the cumulative distribution (cdf) of $\lambda_{\rm max}$
is a product of all the $h_n$'s for $0\leq n<N$. Our expansion will thus give us the behaviour of the cdf of $\lambda_{\rm max}$
for large $N$ only if we can show that it is valid not only for fixed $n$ but also for $n$ of 
order up to $N$.
This constraint of validity will be discussed later. We will see that this expansion is actually valid only on the
right of the mean value, i.e. for $y>\sqrt{\frac{2 N}{\alpha}}$ or equivalently $z>\sqrt{\frac{2 }{\alpha}}$.
This method allows us to describe the right tail of the large deviation of the distribution of 
$\lambda_{ \rm max}$, i.e.
$\mathcal{P}(\lambda_{\rm max}= t)$ with $t>\sqrt{\frac{2  N}{\alpha}}$ and $\left|t-\sqrt{\frac{2 N}{\alpha}}\right|
\approx O(\sqrt{N})$.

\subsection{Expansion of $R_n$ and $S_n$}

Let us start by expanding the initial conditions for large $N$.
With the scaling 
$y=z \sqrt{N}$ with $z\approx O(1)$, equations \eqref{eq:CI} become (for $z>0$)
for large $N$:
\begin{eqnarray}\label{eq:CIlargeN}
h_0(z \sqrt{N},\alpha)&\approx& \sqrt{\frac{\pi}{\alpha}}-\frac{1}{2 z \alpha \sqrt{N} }\:  e^{-N \alpha z^2}
+...
\nonumber\\
 S_0(z \sqrt{N},\alpha)&\approx&-\frac{1}{2 \sqrt{\alpha \pi}} e^{-N \alpha z^2} \nonumber\\
R_1(z \sqrt{N},\alpha)
&\approx& \frac{1}{2 \alpha}-\frac{z \sqrt{N}}{2 \sqrt{\alpha \pi}} e^{-N \alpha z^2}
\end{eqnarray}

The dominant term for large $N$ corresponds to the limit $y\rightarrow \infty$
(see previous section):
$\sqrt{\frac{\pi}{\alpha}}=\int_{-\infty}^{+\infty} d\lambda\, e^{-\alpha \lambda^2}
=h_0(\infty,\alpha)$.
Therefore, ignoring the exponentially small correction for large $N$ leads
to $R_n(z \sqrt{N},\alpha)\approx R_n(\infty,\alpha)=\frac{n}{2 \alpha}$
and $ S_n(z \sqrt{N},\alpha)\approx 0$.

We want to determine the first correction for large $N$.
Let us make the following ansatz:
\begin{equation}\label{eq:ansatzRn}
R_n(z \sqrt{N},\alpha) \approx \frac{n}{2 \alpha} +
c_n\;  e^{-N \alpha z^2}\;\;
{\rm and} \;\; S_n(z \sqrt{N},\alpha) \approx d_n\;  e^{-N \alpha z^2}\;\;\;\;\;\;
\end{equation}
where $c_n=c_n(z \sqrt{N},\alpha)$ and $d_n=d_n(z \sqrt{N},\alpha)$
are expected to be polynomials of $z \sqrt{N}$.
This will be valid as long as 
$c_n(y,\alpha) e^{-N \alpha z^2}\ll \frac{n}{2 \alpha}$ where $y=z\sqrt{N}$. 

The initial conditions in Eq. \eqref{eq:CIlargeN} give:
\begin{equation}\label{eq:c1}
c_0(y,\alpha)=0\;, \;\; c_1(y,\alpha)=-\frac{y}{2 \sqrt{\alpha \pi}}\;\;
{\rm and}\;\;
 d_0(y,\alpha)=-\frac{1}{2 \sqrt{\alpha \pi}}
\end{equation}

Let us
 replace $R_n$ and $S_n$ in the recursion equation \eqref{eq:daRn} and \eqref{eq:dyRn} by the ansatz in Eq. \eqref{eq:ansatzRn}.
We see that $S_n^2$ and $S_{n-1}^2$ are actually negligible in the equation \eqref{eq:daRn} giving $R_{n+1}$, as they are exponentially smaller
than the $R_k$'s.
We thus get  recursion relations for the coefficients $c_n(y,\alpha)$ and $d_n(y,\alpha)$:
\begin{eqnarray}\label{eq:RecCn}
c_{n+1}-c_{n-1}&=&\frac{2}{n} \left\{ -\frac{\partial \left( \alpha c_n \right)}{\partial\alpha}
+\alpha y^2 c_n
\right\}\nonumber\\
d_n-d_{n-1}&=&\left(2 \alpha y c_n - \frac{\partial c_n}{\partial y}\right) \frac{1}{n}
\end{eqnarray}
$h_n$ is the product of the $R_k$, it can thus be expressed in terms of $c_k$.
As $Z_N$ 
is a function of the $h_n$'s, it is thus a function of $h_0$ and  the $R_n$'s for $ 1\leq n <N$, we can use the ansatz \eqref{eq:ansatzRn} 
only if
$c_n(z \sqrt{N},\alpha) e^{-N \alpha z^2}\ll \frac{n}{2 \alpha}$ for all $n<N$.
In this case we can write:
\begin{eqnarray}
\ln h_n(y,\alpha)&=&\ln h_0+\sum_{k=1}^n \ln R_k
\approx \ln h_0 +\sum_{k=1}^n \left[  \ln\left(\frac{k}{2 \alpha}\right) + \frac{2 \alpha c_k}{k} \: e^{-N \alpha z^2 }
 \right]\nonumber \\
&\approx& \ln h_n(\infty,\alpha) +\left[ -\frac{1}{2 y \sqrt{\pi }}
+\sum_{k=1}^n  \frac{2 \alpha c_k(y,\alpha)}{k}
\right]\: 
 e^{-N \alpha z^2 }
\end{eqnarray}
where $y=z \sqrt{N}$, and $\ln Z_N(y,\alpha) = \sum_{n=0}^{N-1} \ln h_n(y,\alpha) $
is thus given by
\begin{equation}\label{eq:Znc}
 \ln Z_N(y,\alpha)\approx
\ln Z_N(\infty,\alpha) + \left[ -\frac{N}{2 y \sqrt{\pi }}
+\sum_{n=0}^{N-1}\sum_{k=1}^n  \frac{2 \alpha c_k(y,\alpha)}{k}
\right]\: 
 e^{-N \alpha z^2 } 
\end{equation}
The partition function only depends on the $c_k$'s.
We want now to solve the recursion relation for the $c_k$'s in Eq. \eqref{eq:RecCn}.
We do not need to determine the $d_k$'s.

\subsection{Solution of the recursion for the $c_n$}

Let us define $\xi$ and $G_n$ such that:
\begin{equation}\label{eq:defXiGn}
\xi=\alpha y^2 =N \alpha z^2\;\;\;\;{\rm and}
\;\;\;\;  c_n(y,\alpha)= -\frac{y^2}{2 \sqrt{\pi \xi}} \: G_n(\xi) \:
\end{equation}
$\xi$ is large for large $N$ (proportional to $N$).
$G_n(\xi)$ depends only on $\xi=\alpha y^2$.
This can easily be shown by recurrence with initial condition
$G_1(\xi)=1$ (as $c_1$ is given by Eq. \eqref{eq:c1}).
The recursion \eqref{eq:RecCn} for the $c_n$'s becomes
\begin{equation}\label{eq:RecGn}
G_{n+1}(\xi)-G_{n-1}(\xi)=\frac{2}{n}\left\{ \left(\xi-\frac{1}{2}\right) G_n(\xi) +\xi \: G_n'(\xi) \right\}
\end{equation}
with initial condition $G_0(\xi)=0$ and $G_1(\xi)=1$.
By recurrence again, it is easy to show that $G_n(\xi)$ is a polynomial of $\xi$
of degree $(n-1)$, with leading coefficient $\frac{2^{n-1}}{(n-1)!}$.

Let us consider the generating function of the $\left\{ G_n(\xi)\right\}$:
\begin{equation}\label{eq:defF}
F(\xi,x)=\sum_{n=1}^{\infty} x^n \, G_n(\xi)
\end{equation}
The $G_n(\xi)$ are obtained from $F$ by a contour integration:
\begin{equation}\label{eq:Gncontour}
G_n(\xi)=\oint_{\mathcal{C}} \frac{d x}{2 i \pi}\frac{1}{x^{n+1}}\, F(\xi,x)
\end{equation}
where $C$ is a contour in the complex plane that encircles the origin $x=0$
in such a way that all singularities of $F(\xi,x)$ (as a function of $x$ for fixed $\xi$)
are outside the contour.

From Eq. \eqref{eq:RecGn} and the definition of $F$, we deduce that
$F(\xi,x)$ satisfies the following  partial differential equation:
\begin{equation}
(1-x^2) \frac{\partial F}{\partial x}+2 \xi \frac{\partial F}{\partial \xi}=
\left[ x+\frac{1}{x}+2 \xi -1 \right] F
\end{equation}
This equation together with the  requirement that
$F(\xi,x)\approx x+O(x^2)$ as $x\rightarrow 0$ (as $G_1=1$) determines uniquely $F(\xi,x)$.
We find:
\begin{equation}
F(\xi,x)=\frac{x}{(1+x)\sqrt{1-x^2}}\, e^{\frac{2 \xi x}{x+1}}
\end{equation}
$G_n(\xi)$ is given by the contour integral in Eq. \eqref{eq:Gncontour}
where the contour $C$  encircles $x=0$ in such a way that 
$x=1$ and $x=-1$ are outside of the contour.

Let us compute $G_n(\xi)$ with $\xi=N \alpha z^2$ for 
large $N$, fixed $z$ and $n=c N$ with fixed $0<c\leq 1$. We have
\begin{equation}
G_n(\xi)=\oint_{\mathcal{C}} \frac{d x}{2 i \pi}\frac{1}{x^{n+1}}\, F(\xi,x)
=\oint_{\mathcal{C}} \frac{d x}{2 i \pi} \frac{e^{N\Phi_c(x)}}{(1+x)\sqrt{1-x^2}}
\end{equation}
where 
\begin{equation}
\Phi_c(x)=\frac{2 \alpha z^2 x}{x+1}-c\, \ln x\;\; {\rm with }\;\; c=\frac{n}{N}
\label{phic}
\end{equation}
is of order one for large $N$ when $n=c N$ with $c$ of order one.
For fixed $c\leq 1$ and for large $N$, the contour integral can thus be computed
using a saddle point method.
The integral will be dominated by the neighbourhood of $x^*$ such that:
\begin{equation}
\frac{d\Phi_c}{d x}\Big|_{x^*}=0\;\;\;{\rm ie}
\;\;\; \frac{2 \alpha z^2}{(1+x^*)^2}=\frac{n}{N x^*}=\frac{c}{x^*}
\end{equation}
There exists a real solution for $x^*$ iff
$z^2 > \frac{2 c}{\alpha}$ ie $z^2 > \frac{2 n}{\alpha N}$.
We want this condition to be satisfied for all $n<N$, therefore
we must have $z^2>\frac{2}{\alpha}$, i.e. $z> \sqrt{\frac{2}{\alpha}}$.
Our method can only describe the regime $z> \sqrt{\frac{2}{\alpha}}$,
ie $y>\sqrt{\frac{2 N}{\alpha}}$, which corresponds to the right tail of the distribution
${\rm P}(\lambda_{\rm max} \leq y)$ (region where $\lambda_{\rm max}$ is above its mean value).
Let us call the critical point $y_{\rm cr}=\sqrt{\frac{2 N}{\alpha}}$, i.e.
$z_{\rm cr}=\sqrt{\frac{2}{\alpha}}$.

For $y>y_{\rm cr}$, there are two real solutions
$x^*=-1+\frac{N \alpha z^2 }{n}\left[ 1\pm \sqrt{1-\frac{2 n}{N \alpha z^2} } \right]$.
The contour $C$ must encircle $0$ but not $1$ and $-1$, therefore
we impose $-1<x^*<1$.
This implies
\begin{eqnarray}\label{eq:xstar}
x^*&=&-1+\frac{N \alpha z^2 }{n}\left[ 1 - \sqrt{1-\frac{2 n}{N \alpha z^2} } \right]
=\frac{N \alpha z^2}{2 n}\left[ 1-\sqrt{1-\frac{2 n}{N \alpha z^2}} \right]^2 \nonumber \\
&=&-1+\frac{\xi }{n}\left[ 1 - \sqrt{1-\frac{2 n}{\xi} } \right]
=\frac{\xi}{2 n}\left[ 1-\sqrt{1-\frac{2 n}{\xi}} \right]^2
\end{eqnarray}
Thus
\begin{equation}
 \Phi_c(x^*)=\alpha z^2 \left[ 1-\sqrt{1-\frac{2 c}{ \alpha z^2}} \right]
-2  c \ln\left\{\sqrt{\frac{ \alpha z^2}{2 c}}-\sqrt{\frac{ \alpha z^2}{2 c}-1}\right\}
\end{equation}
The saddle point gives for large $N$:
\begin{equation}\label{eq:SolGn}
G_n(\xi)\approx \frac{1}{2 \pi}
\frac{e^{N\Phi_c(x^*)}}{(1+x^*)\sqrt{1-{x^*}^2}} \sqrt{\frac{2 \pi}{N 
\left|\frac{d^2 \Phi_c}{d x^2}\big|_{x^*}\right|}}
\end{equation}
where
\begin{equation*}
\left|\frac{d^2 \Phi_c}{d x^2}\big|_{x^*}\right|= \frac{4\: \sqrt{\frac{\alpha
  z^2}{c}-2} \; }{\left( \sqrt{\frac{\alpha z^2}{
    c}-2}
-\sqrt{ \frac{\alpha z^2}{
    c}} \right)^4\, \sqrt{\frac{\alpha z^2}{c}}}
\end{equation*}
In this subsection we have found, as given in Eq. \eqref{eq:SolGn},  the expression of $G_n(\xi)$ and thus 
the solution $c_n(y,\alpha)= -\frac{y^2}{2 \sqrt{\pi \xi}} \: G_n(\xi) $ (with $\xi = N \alpha z^2=\alpha y^2$)
of the recursion relation \eqref{eq:RecCn} for large $N$
and $n=c N$ with fixed $0<c\leq 1$.
We have also shown that the validity of our approximation
is the regime $y>y_{\rm cr}$ with $y_{\rm cr}=\sqrt{\frac{2 N}{\alpha}}=\langle \lambda_{\rm max}\rangle$.

\subsection{Computation of the distribution of $\lambda_{\rm max}$ for large $N$}

We want to compute  for large $N$ and for $y>y_{\rm cr}$ (with the scaling $y=z\sqrt{N}$ for
large $N$) the cdf
$\mathbb{P}_N\left(\lambda_{\rm max} \leq y\right) =\frac{Z_N \left(y,\alpha\right)}{Z_N(\infty,\alpha)}$.
 Using Eq. \eqref{eq:Znc} and the definition of $G_n$ in Eq. \eqref{eq:defXiGn}, we get:
\begin{eqnarray}
\ln\mathbb{P}_N\left(\lambda_{\rm max} \leq y\right)&=&
\ln Z_N(y,\alpha) - \ln Z_N(\infty,\alpha)\nonumber
\\
&\approx&
-\frac{e^{-\xi}}{2 \sqrt{\pi \xi}} \left\{ 
1+2 \xi \sum_{n=1}^{N-1} \left(\frac{N-n}{n}\right) G_n(\xi)
\right\}
\end{eqnarray}
Therefore we need to compute
$I_N(\xi)\equiv \sum_{n=1}^{N-1} \left(\frac{N-n}{n}\right) G_n(\xi)$ for large $N$ and
$\xi=N \alpha z^2$ (with fixed $\alpha$ and $z$).
For that purpose, we do not actually need to use the approximate expression of 
$G_n$ for $n$ of order $N$ that we derived in the previous subsection.
We can use the formal expression of
$G_n$ as a contour integral
$G_n(\xi)=\oint_{\mathcal{C}} \frac{d x}{2 i \pi}\frac{1}{x^{n+1}}\, F(\xi,x)$
and compute the sum over $n$ before computing the integral by saddle point method.
In particular we have:
\begin{equation*}
\sum_{n=1}^{N-1} G_n(\xi)=\oint_{\mathcal{C}} \frac{d x}{2 i \pi}\frac{F(\xi,x)}{x (x-1)}+
\oint_{\mathcal{C}} \frac{d x}{2 i \pi}\frac{F(\xi,x)}{x^N (1-x)}
=
\oint_{\mathcal{C}} \frac{d x}{2 i \pi}\frac{F(\xi,x)}{x^N (1-x)}
\end{equation*}
The function $\frac{F(\xi,x)}{x (x-1)}=\frac{1}{(x^2 -1)\sqrt{1-x^2}}\, e^{\frac{2 \xi x}{x+1}}$
has indeed no singularity at the origin,  its integral is thus zero.
On the other hand, we have:
\begin{equation*}
\sum_{n=1}^{N-1} \frac{G_n(\xi)}{n}=\oint_{\mathcal{C}} \frac{d x}{2 i \pi}\left(\sum_{n=1}^{N-1}\frac{1}{n x^{n+1}}\right)\, F(\xi,x)
=\oint_{\mathcal{C}} \frac{d x}{2 i \pi}
\frac{F(\xi,x)}{ x^N} \frac{\,_2F_1\left( 1,1-N,2-N,x \right)}{N-1}
\end{equation*}
where $\,_2F_1$ is a hypergeometric function:
$\,_2F_1(a,b,c,z)=\sum_{k\geq 0} \frac{(a)_k (b)_k}{(c)_k} \frac{z^k}{k!}$
with $(a)_k=a (a+1)...(a+k-1)$.
For large $N$, we find:
\begin{eqnarray*}
\frac{ \,_2F_1\left( 1,1-N,2-N,x \right)}{N-1}=\sum_{k\geq 0} \frac{x^k}{-1-k+N} 
\approx
\sum_{k\geq 0}  x^k \left(\frac{1}{N}+\frac{1-k}{N^2}+...\right)
\approx \frac{1}{N (1-x)} +\frac{1}{(1-x)^2 N^2}+...
\end{eqnarray*}
Therefore we get for large $N$
\begin{equation}
I_N(\xi)=\sum_{n=1}^{N-1} \left(\frac{N-n}{n}\right) G_n(\xi)\approx
\frac{1}{N}\oint_{\mathcal{C}} \frac{d x}{2 i \pi}
\frac{F(\xi,x)}{ x^N (1-x)^2}+...
\end{equation}
Equivalently we can write:
\begin{equation}
I_N(\xi) \approx \frac{1}{N}\oint_{\mathcal{C}} \frac{d x}{2 i \pi}
\frac{x}{(1+x)^{\frac{3}{2}} (1-x)^{\frac{5}{2}}}\: e^{N \Phi_1(x)}
\end{equation}
where $\Phi_1(x)=\frac{2 \alpha z^2 x}{x+1} - \ln x$
(see \eqref{phic}).
For large $N$, the saddle point method thus gives
\begin{equation}
I_N(\xi) \approx \frac{1}{N} \frac{1}{2  \pi}
\frac{x^*}{(1+x^*)^{\frac{3}{2}} (1-x^*)^{\frac{5}{2}}}\: e^{N \Phi_1(x^*)}\:
\sqrt{\frac{2 \pi}{N 
\left|\frac{d^2 \Phi_c}{d x^2}\big|_{x^*}\right|}}
\end{equation}
where $x^*$ is given in Eq. \eqref{eq:xstar}
 with $n=N$:
\begin{equation}
x^*=-1+ \alpha z^2 \left[ 1 - \sqrt{1-\frac{2 }{ \alpha z^2} } \right]
=\frac{ \alpha z^2}{2 }\left[ 1-\sqrt{1-\frac{2 }{ \alpha z^2}} \right]^2
\end{equation}
Thus we find
\begin{equation}
I_N(\xi) \approx \frac{1}{N^{\frac{3}{2}}} \frac{1}{\sqrt{2  \pi}}
\frac{1}{4 \sqrt{\alpha z^2} (\alpha z^2-2)^{\frac{3}{2}}}\: e^{N \Phi_1(x^*)}
\end{equation}
with
\begin{equation}
 \Phi_1(x^*)=\alpha z^2 \left[ 1-\sqrt{1-\frac{2 }{ \alpha z^2}} \right]
-2   \ln\left\{\sqrt{\frac{ \alpha z^2}{2 }}-\sqrt{\frac{ \alpha z^2}{2 }-1}\right\}
\end{equation}
Therefore
\begin{eqnarray*}
\ln\mathbb{P}_N\left(\lambda_{\rm max} \leq y\right)&=&
\ln Z_N(y,\alpha) - \ln Z_N(\infty,\alpha)\nonumber \\
&\approx&
-\frac{e^{-\xi}}{2 \sqrt{\pi \xi}} \left\{ 
1+2 \xi  I_N(\xi)
\right\}\\
&\approx& 
-\frac{e^{-N \alpha z^2}}{2 \sqrt{\pi N \alpha z^2}} \left\{ 
1+ \, \frac{\sqrt{\alpha z^2}}{\sqrt{N}2 \sqrt{2  \pi}
 (\alpha z^2-2)^{\frac{3}{2}}}\: e^{N \Phi_1(x^*)}
\right\}
\end{eqnarray*}
As $\Phi_1(x^*) > 0$ for $z>z_{\rm cr}$, i.e. $\alpha z^2>2$, the first term
in the parenthesis can be neglected for large $N$:
\begin{equation}
\ln\mathbb{P}_N\left(\lambda_{\rm max} \leq y\right)
\approx
-\frac{e^{-N \alpha z^2}}{4 \pi N } \:
 \frac{1}{ \sqrt{2  }
 (\alpha z^2-2)^{\frac{3}{2}}}\: e^{N \Phi_1(x^*)}
\end{equation}
with $y=z \sqrt{N}$.
Therefore we get the expression of the right tail of the cdf of $\lambda_{\rm max}$:
\begin{equation}
\ln\mathbb{P}_N\left(\lambda_{\rm max} \leq z \sqrt{N}\right)
\approx -\frac{1}{4 \pi N \sqrt{2} (\alpha z^2-2)^{\frac{3}{2}}} \: e^{-2 N \psi_+(z)}\;\;
{\rm for}\;\; z>\sqrt{\frac{2}{\alpha}}\;\;\;\;\;\:\;\;\;
\end{equation}
where the rate function $\psi_+(z)=\frac{\alpha z^2-\Phi_1(x^*)}{2}$ is given by:
\begin{equation}
\psi_+(z)=\frac{\alpha z^2}{2} \left[ \sqrt{1-\frac{2 }{ \alpha z^2}} \right]
+  \ln\left\{\sqrt{\frac{ \alpha z^2}{2 }}-\sqrt{\frac{ \alpha z^2}{2 }-1}\right\}
\label{psi+z}
\end{equation}
We have thus found 
\begin{equation}\label{eq:RightTail}
\mathbb{P}_N\left(\lambda_{\rm max} \leq z \sqrt{N}\right)
\approx 1-\frac{1}{4 \pi N \sqrt{2} (\alpha z^2-2)^{\frac{3}{2}}} \: e^{-2 N \psi_+(z)}\;\;
{\rm for}\;\; z>\sqrt{\frac{2}{\alpha}}\;\;\;\;\;\:\;\;\;
\end{equation}
Deriving with respect to $t=z \sqrt{N}$  we get
an equivalent of the probability density function of $\lambda_{\rm max}$ for large $N$:
\begin{equation}\label{eq:pdfEquiv}
\mathcal{P}\left(\lambda_{\rm max} =t \right)
\approx \frac{\sqrt{\alpha N}}{ 2 \pi \sqrt{2 }\, (\alpha t^2-2 N)} \: e^{-2 N \psi_+\left(\frac{t}{\sqrt{N}}\right)}\;\;
{\rm for}\;\; t>\sqrt{\frac{2 N}{\alpha}}\;\;,\;\;
\left|t-\sqrt{\frac{2 N}{\alpha}}\right| \approx O(\sqrt{N})\;\:\;\;\;
\end{equation}
We thus recover the dominant order for large $N$ (given by $\psi_+(z)$) that was derived in
\cite{vergassolaSM} by a Coloumb gas method, but in addition this method also
provides us with the first correction term to the dominant order. 
\\

Let us now see how this precise right tail large deviation behavior in \eqref{eq:RightTail}
behaves when $z \to \sqrt{\frac{2}{\alpha}}$ from the right. Using the leading expansion of 
$\psi_{+}(z)$ around $z=\sqrt{2}$ in \eqref{rightv1} and setting  
$y=\sqrt{\frac{2 N}{\alpha}}+N^{-1/6} \frac{x}{\sqrt{2 \alpha}}$, i.e., 
$z=y/\sqrt{N}=\sqrt{\frac{2 }{\alpha}}+N^{-2/3} \frac{x}{\sqrt{2 \alpha}}$
one finds from \eqref{eq:RightTail}
\begin{equation}
\label{ldrtail}
\mathbb{P}_N\left(\lambda_{\rm max} \leq y \right)\approx 1- \frac{1}{16 \pi x^{\frac{3}{2}}}\,
 e^{-\frac{4 }{3} x^{\frac{3}{2}}}.
\end{equation}
On the other hand, using the boundary condition $q(z)\approx {\rm Ai}(z)\approx 
\frac{1}{2 \sqrt{\pi}z^{1/4}}\, e^{-\frac{2}{3}z^{3/2}}$ as
$z\to \infty$ in the definition of the Tracy-Widom function \eqref{f2tw}, one can easily derive the
precise leading asymptotics of its right tail,  
$F_2(x)\approx 1- \frac{1}{16 \pi x^{\frac{3}{2}}}\,
 e^{-\frac{4 }{3} x^{\frac{3}{2}}}$ as $x\rightarrow  \infty$. Thus, our right large deviation
function for small argument (when the fluctuation of $\lambda_{\rm max}$ to the right of its 
mean value $\sqrt{2N/\alpha}$ is of $O(N^{-1/6})$) in \eqref{ldrtail}, matches smoothly with the 
precise right tail of the Tracy-Widom
distribution $F_2(x)$.

\section{Double scaling limit and Tracy-Widom distribution}
\label{sec:TW}

In this section, we provide an elementary derivation of the Tracy-Widom law for the GUE based on 
simple scaling analysis of the recursion relations derived in Section 2 in the vicinity
of the critical point $y=y_{\rm cr}= \sqrt{\frac{2N}{\alpha}}$. This derivation, in our opinion,
is mathematically simpler than the original derivation by Tracy and Widom~\cite{TW1,TW2} as
it avoids the sophisticated asymptotic analysis of Fredholm determinants.
The derivation of the Painlev\'e II equation from the scaling analysis of recursion
relations that we follow here is similar in spirit (though rather different in details)
to the analysis of the partition function in the two dimensional Yang-Mills theory
on a sphere by Gross and Matytsin~\cite{gross}.   

Let us recall that
the Tracy-Widom distribution $F_2(x)$ is defined as
\begin{equation}\label{eq:defTW}
 F_2(x)=\exp\left\{
-\int_{x}^{\infty}ds (s-x) q^2(s)
\right\}
\end{equation}
where $q(x)$ satisfies the Painlev\'e II equation with the boundary condition
\begin{eqnarray}
\label{eq:Painleve}
 q''(x)&=&2 q^3(x) + x q(x)\;\;\;\;\;\; \textrm{Painlev\'e II}\nonumber \\
q(x) &\approx& \frac{1}{2 \sqrt{\pi}x^{1/4}}\, e^{-\frac{2}{3}x^{3/2}}
\;\;\;\; {\rm as}\;\; x\rightarrow \infty
\end{eqnarray}
From Eq. \eqref{eq:defTW}, it follows that
$\frac{d^2 \ln F_2(x)}{dx^2}=-q^2(x)$.
\\

We want to show that for large $N$ the probabality of small typical fluctuations of 
$\lambda_{\rm max}$ around its mean value $\sqrt{\frac{2 N}{\alpha}}$ are described 
by the Tracy-Widom distribution. For this, we need to first estimate how do these
typical fluctuations scale with $N$ for large $N$. In the vicinity of the mean 
$\sqrt{\frac{2N}{\alpha}}$, let us then write
\begin{equation}
\label{typ_fluc1}
\lambda_{\rm max}\approx\sqrt{\frac{2 N}{\alpha}}+\frac{1}{\sqrt{2\alpha}}\,N^{\gamma}\, x
\end{equation}
where $N^{\gamma}$ is the scale of the typical fluctuation and the random variable $x$
has an $N$ independent distribution for large $N$. Evidently the exponent $\gamma<1/2$
(so that the fluctuation is less than the mean) whose precise value is yet to be determined.
Note also that since $\lambda_{\rm max}$ always appears in the distribution 
$\mathbb{P}_N(\lambda_{\rm max}\le y)$ in the scaling 
combination $y\sqrt{\alpha}$ (see Eq. \eqref{rescaled1}), we have chosen the 
prefactor of the fluctuation term 
as $1/\sqrt{2\alpha}$ which then ensures that the random variable $x$ describing 
the typical fluctuation is also independent of $\alpha$.

One way to estimate the exponent $\gamma$ is from the right large deviation tail
computed in \eqref{eq:RightTail} in the previous section. The right tail
in \eqref{eq:RightTail} describes the probability of large fluctuations of
$O(\sqrt{N})$ to the right of the mean. Assuming that the right tail
behaviour continues to hold even for fluctuations smaller than $\sqrt{N}$,
we substitute 
$z\sqrt{N}= \sqrt{\frac{2 N}{\alpha}}+ \frac{1}{\sqrt{2\alpha}}\,N^{\gamma}\, x$
in \eqref{eq:RightTail}. This gives
\begin{equation}
\mathbb{P}_N\left(\lambda_{\rm max} \leq \sqrt{\frac{2 N}{\alpha}}+ 
\frac{1}{\sqrt{2\alpha}}\,N^{\gamma}\, x\right)\approx 1- 
\frac{1}{N^{\frac{1}{4}+\frac{3 \gamma}{2}}}\frac{1}{16 \pi  x^{\frac{3}{2}}} \: 
e^{-\frac{4}{3} x^{3/2} N^{\frac{3 \gamma}{2}+\frac{1}{4}}
}
\label{gamma1}
\end{equation}
valid for $x>0$, $x$ large. Assuming that this continues to hold even for 
not so large $x$ (so that it even captures the tail of the distribution of 
typical small fluctuations), we expect that in terms of this rescaled variable $x$,
the tail of the distribution in \eqref{gamma1} is independent of $N$ for large $N$.
Clearly, for this to happen the power of $N$ must be zero both inside
the exponential as well as in the prefactor in \eqref{gamma1}, indicating
that $\frac{1}{4}+\frac{3 \gamma}{2}=0$, thus $\gamma=-\frac{1}{6}$.
Hence, the correct scaling describing typical fluctuations, for large $N$, is 
given by
\begin{equation}
\label{eq:scalingTW}
\lambda_{\rm max}=\sqrt{\frac{2 N}{\alpha}} +{\frac{1}{\sqrt{2\alpha}}}\, N^{-\frac{1}{6}}\, x
\end{equation}
where $x$ has an $N$ independent distribution that we now have to compute and show
that it is given by the Tracy-Widom function $F_2(x)$.

The meaning of {\em double scaling limit} is now clear. It simply says the following.
Consider the cdf $\mathbb{P}_N\left(\lambda_{\rm max} \leq y\right)$ or rather
its logarithm (for convenience) $\ln \mathbb{P}_N\left(\lambda_{\rm max} \leq y\right)$.
In general, it is a function of two variables $y$ and $N$. However, in the vicinity
of the mean
$y\to \sqrt{\frac{2N}{\alpha}}$, if one takes the limit $y-\sqrt{\frac{2N}{\alpha}} \to 0$ and
$N\to \infty$,
but keeping the scaling combination $x=\sqrt{2\alpha}\, N^{1/6}\, (y-\sqrt{\frac{2N}{\alpha}})$
fixed, this function of two variables collapses into a function of the single scaled variable
$x$
\begin{equation}
\ln \mathbb{P}_N\left(\lambda_{\rm max} \leq y\right) \to f\left(\sqrt{2\alpha}\, N^{1/6}\,
\left(y-\sqrt{\frac{2N}{\alpha}}\right)\right)
\label{dscale1}
\end{equation}
and our job is to to show that this scaling function $f(x)= \ln F_2(x)$ where
$F_2(x)$ is the Tracy-Widom function defined in \eqref{eq:defTW}. In other
words, we want to show that $f''(x)=-q^2(x)$ where $q(x)$ satisfies
the Painlev\'e II equation \eqref{eq:Painleve}.

Our starting point is the definition of the cdf in \eqref{eq:CdfLmax}. 
From Eq. \eqref{eq:Znh} it is easy to see
that the partition function $Z_N$ satisfies the recursion
\begin{equation}
\label{pfrecur1}
 \frac{Z_{N-1}(y,\alpha)\, Z_{N+1}(y,\alpha)}{Z_N^2(y,\alpha)}=\frac{h_N(y,\alpha)}{h_{N-1}(y,\alpha)}
=R_N(y,\alpha)
\end{equation}
Taking logarithm and using the definition in \eqref{eq:CdfLmax} we get
\begin{equation}\label{eq:PnRn}
\ln \mathbb{P}_{N+1}\left(\lambda_{\rm max}\leq y\right)+\ln\mathbb{P}_{N-1}\left(\lambda_{\rm max}\leq y\right)
-2 \ln\mathbb{P}_{N}\left(\lambda_{\rm max}\leq y\right)=\ln\left(\frac{R_N(y,\alpha)}{R_N(\infty,\alpha)}\right)
\end{equation}
In the double scaling 
limit, we will now substitute the anticipated scaling form in \eqref{dscale1} 
for the logarithm of the cdf on the left hand side of \eqref{eq:PnRn}. But we need
to first evaluate $\ln \mathbb{P}_{N\pm 1}\left(\lambda_{\rm max}\leq y\right)$.
Replacing $N$ by $N\pm 1$ in \eqref{dscale1} and expanding for large $N$, with
$x= \sqrt{2\alpha}\, N^{1/6}\,
\left(y-\sqrt{\frac{2N}{\alpha}}\right)$ fixed, we get
\begin{eqnarray}
\ln \mathbb{P}_{N\pm 1}\left(\lambda_{\rm max}\leq y\right)& =& f\left(\sqrt{2\alpha}\, 
\left(N\pm 1\right)^{1/6}\,
\left(y-\sqrt{\frac{2(N\pm 1)}{\alpha}}\right)\right) \nonumber \\
&=& f\left(x\mp N^{-1/3}\pm \frac{x}{6N}\pm \frac{N^{-4/3}}{12}+\dots\right) \nonumber \\
&=& f(x) \mp N^{-1/3} f'(x) + \frac{N^{-2/3}}{2}\, f''(x) + O(N^{-1}).
\label{scalingexpan1}
\end{eqnarray}
Substituting this result in \eqref{eq:PnRn} we get for the left hand side 
\begin{eqnarray}\label{eq:PnOrd2}
\ln\mathbb{P}_{N+1}\left(\lambda_{\rm max}\leq y\right)+\ln\mathbb{P}_{N-1}\left(\lambda_{\rm max}\leq y\right)
-2 \ln\mathbb{P}_{N}\left(\lambda_{\rm max}\leq y\right)\nonumber\\\approx 
N^{-2/3} \, f''(x)+O(N^{-1})\;\;\;\;\;\;\;\;\;\;\;
\end{eqnarray}
From Eq. \eqref{eq:PnRn} and \eqref{eq:PnOrd2}, we get for large $N$
\begin{equation}
N^{-2/3} \, f''(x)\approx \ln\left(\frac{R_N(y,\alpha)}{R_N(\infty,\alpha)}\right)
\approx \ln\left(\frac{R_N(y,\alpha)}{N/(2 \alpha)}\right)
\end{equation}
as $R_N(\infty,\alpha)=N/(2 \alpha)$
(see Eq. \eqref{eq:SnRnNorm}).
This suggests that in this scaling limit, $R_N$ must scale as
$ R_N\left(y,\alpha\right)\approx 
\frac{N}{2 \alpha}\left(1+N^{-\frac{2}{3}}\: f''(x)+...\right)$.
More precisely, this leads us to the following large $N$ expansion
of $R_N(y, \alpha)$ in the double scaling limit 
\begin{equation}\label{Rnapprox}
 R_N\left(y,\alpha\right)\approx \frac{N}{2 \alpha}\left(1+N^{-\frac{2}{3}}\: r_1(x)+
N^{-1} r_2(x)+N^{-\frac{4}{3}} r_3(x)+...\right),
\end{equation}
where
\begin{equation}
f''(x)=r_1(x)
\label{fxr1x}
\end{equation}
and $r_2(x)$, $r_3(x)$ etc. describing the higher order scaling corrections.
Thus, if we can now determine the first subleading scaling function
$r_1(x)$ in the expansion of $R_N\left(y,\alpha\right)$, then we can determine
$f(x)$ by integrating $r_1(x)$ twice. So, our next task is to determine $r_1(x)$
by analysing the recursion relations \eqref{eq:daRn} and \eqref{eq:dyRn} (setting $n=N$) in the
double scaling limit.

We now know, from \eqref{Rnapprox}, how $R_N(y,\alpha)$ behaves in the scaling limit with the
scaling combination $x=\sqrt{2\alpha}\,N^{1/6}\,\left(y-\sqrt{\frac{2N}{\alpha}}\right)$ fixed.
In order to analyse the recursion relations \eqref{eq:daRn} and \eqref{eq:dyRn}, we also need 
to know how $S_N(y,\alpha)$ behaves in this scaling limit. In order to match the 
leading $N$ behavior of $R_N(y,\alpha)$ with $x$ fixed in \eqref{eq:dyRn}, it is not difficult
to see that to leading order for large $N$, $S_N(y,\alpha)$ must have the following scaling 
behaviour
\begin{equation}
S_N(y,\alpha)
\approx\frac{N^{-1/6}}{\sqrt{2 \alpha}}\,s_1\left(\sqrt{2\alpha}\,N^{1/6}
\left(y-\sqrt{\frac{2N}{\alpha}}\right)\right)+ O(N^{-1/2})
\label{snscaling1}
\end{equation}
where $s_1(x)$ is the leading order scaling function. Let us first evaluate the difference
$S_{N-1}(y,\alpha)-S_N(y,\alpha)$ that appears in \eqref{eq:dyRn}. Replacing
$N$ by $N-1$ in \eqref{snscaling1}, expanding for large $N$, we get 
\begin{equation}
S_{N-1}(y,\alpha)-S_N(y,\alpha)\approx \frac{N^{-1/2}}{\sqrt{2\alpha}} s_1'(x) + O(N^{-5/6}).
\label{snscaling2}
\end{equation}
It rests to evaluate the partial 
derivative $\frac{\partial \ln R_N(y,\alpha)}{\partial y}$ in \eqref{eq:dyRn}.
From the definition of the scaling variable $x=\sqrt{2\alpha}\,N^{1/6}\,
\left(y-\sqrt{\frac{2N}{\alpha}}\right)$, it follows, using chain rule, 
\begin{eqnarray}
\frac{\partial \ln R_N(y,\alpha )}{\partial y} &=& \frac{\partial \ln R_N(y,\alpha)}{\partial x}
\frac{\partial x}{\partial y} \nonumber \\
&=& \sqrt{2\alpha}\,N^{1/6}\, \frac{\partial \ln R_N(y,\alpha)}{\partial x} \nonumber \\
&=& \sqrt{2\alpha} \, N^{-1/2}\, r_1'(x) + O(N^{-5/6})
\label{snscaling3}
\end{eqnarray}
Finally substituting \eqref{snscaling2} and \eqref{snscaling3} in \eqref{eq:dyRn} (with $n=N$)
we get, 
\begin{equation*}
\sqrt{2 \alpha}\left(r_1'(x)N^{-\frac{1}{2}}+O(N^{-\frac{5}{6}})\right)
\approx \frac{\partial \ln R_N}{\partial y}=2 \alpha (S_{N-1}-S_N)
\approx \sqrt{2 \alpha} \left(N^{-\frac{1}{2}} s_1'(x)+O(N^{-\frac{5}{6}})) \right)
\label{snscaling4}
\end{equation*}
Matching the leading order $N^{-1/2}$ term gives a relation between $s_1(x)$ and $r_1(x)$:
$s_1'(x)=r_1'(x)$, i.e., $s_1(x)=r_1(x)+c_0$ with $c_0$ a constant.
From \eqref{eq:SnRnNorm} and the fact that when $y\to \infty$, $x\to \infty$, it follows
that both the scaling functions $r_1(x)$ and $s_1(x)$ must vanish as $x\to \infty$.
Thus the constant $c_0=0$ and we have, for all $x$,
\begin{equation}
r_1(x)=s_1(x).
\label{r1s1}
\end{equation}

Having determined the relation $r_1(x)=s_1(x)$, we need one more relation between 
these two scaling functions in order to determine them individually.
This will now be done by substituting the scaling solutions for $R_N(y,\alpha)$ (given in 
\eqref{Rnapprox}) and $S_N(y,\alpha)$ (given in \eqref{snscaling1}) into the remaining 
recursion relation \eqref{eq:daRn}.

To analyse \eqref{eq:daRn} (seeting $n=N$), we need to evaluate the derivative 
$\frac{\partial \ln R_N(y,\alpha)}{\partial \alpha}$. 
From the definition of the scaling variable, $x=\sqrt{2\alpha}\,N^{1/6}\,
\left(y-\sqrt{\frac{2N}{\alpha}}\right)$, it follows, that 
$\frac{\partial x}{ \partial \alpha}=\frac{x}{2 \alpha}+\frac{N^{2/3}}{\alpha}$.
We then use the chain rule and \eqref{Rnapprox} to express
\begin{equation*}
\label{rnrecur1}
 \frac{\partial \ln R_N(y,\alpha)}{\partial \alpha}
=\frac{r_1'(x)-1}{\alpha}+\frac{N^{-\frac{1}{3}}}{\alpha}r_2'(x)
+\frac{N^{-\frac{2}{3}}}{2 \alpha}\left[ x r_1'(x)-2 r_1(x) r_1'(x)+2 r_3'(x) \right]+...
\end{equation*}
Again replacing $N$ by $N\pm 1$ in \eqref{Rnapprox} and expanding for large $N$, keeping
$x=\sqrt{2\alpha}\,N^{1/6}\,
\left(y-\sqrt{\frac{2N}{\alpha}}\right)$ fixed, we get
\begin{equation*}
\label{rnrecur2}
R_{N-1}-R_{N+1}\approx \frac{r_1'(x)-1}{\alpha}+\frac{N^{-\frac{1}{3}}}{\alpha} r_2'(x)
+\frac{N^{-\frac{2}{3}}}{6 \alpha}\left[-2 r_1(x)-x r_1'(x)+6 r_3'(x)+r_1'''(x)\right]+...
\end{equation*}
and similarly from \eqref{snscaling1}
\begin{equation*}
\label{snscaling5}
S_{N-1}^2-S_N^2\approx \frac{N^{-\frac{2}{3}}}{\alpha} s_1(x)s_1'(x)+...
\end{equation*}
Substituting these results in \eqref{eq:daRn} and 
matching the 
leading order term ($O(N^{-2/3})$), we get the desired second relation between $r_1(x)$
and $s_1(x)$
\begin{equation*}
\label{r1s12}
x r_1'(x)-2 r_1(x) r_1'(x)=-\frac{2}{3} r_1(x)-
\frac{x}{3} r_1'(x)+\frac{1}{3} r_1'''(x)+2 s_1(x)s_1'(x).
\end{equation*}
Eliminating $s_1(x)$ by using $s_1(x)=r_1(x)$ we finally get a single closed
equation for $r_1(x)$
\begin{equation}\label{eq:r1}
2 x r_1'(x)+r_1(x)=\frac{1}{2} r_1'''(x)+6 r_1(x)r_1'(x).
\end{equation}
Let us write 
\begin{equation}
 r_1(x)=-u^2(x)
\end{equation}
Eq. \eqref{eq:r1} then becomes an equation for $u(x)$:
\begin{equation}
\label{eq:ux}
u(u'''-6 u^2 u'-x u'-u)=-3 u' (u''-x u-2 u^3)
\end{equation}
Let $W(x)=u''(x)-x u(x)-2 u^3(x)$. Then \eqref{eq:ux} becomes
\begin{equation}
\label{eq:Wx}
 u(x) \frac{d W(x)}{dx}=-3 u'(x) W(x)
\end{equation}
which can simply be integrated to give
\begin{equation}
W(x)=\frac{A}{u(x)^3}.
\end{equation}
where $A$ is an arbitrary constant.
Hence we have
\begin{equation}
u''(x)-x u(x)-2 u^3(x)= \frac{A}{u(x)^3}
\label{eq:ux1}
\end{equation}
From the boundary condition $r_1(x)\to 0$ as $x\to \infty$
(which follows from \eqref{eq:SnRnNorm}), it follows using $r_1(x)=u^2(x)$
that $u(x)\to 0$ as $x\to \infty$. Taking $x\to \infty$ in \eqref{eq:ux1} then
fixes the value of the constant $A=0$. Finally, from \eqref{fxr1x}, we have
$f''(x)=r_1(x)=-u^2(x)$ where $u(x)$ satisfies the Painlev\'e II equation
\begin{equation}
u''(x)=x\, u(x) + 2\,u^3(x)
\label{ux}
\end{equation}

To fix the boundary condition for $u(x)$, we again invoke the matching with the right
large deviation tail in \eqref{gamma1}. Taking logarithm of \eqref{gamma1} with
$\gamma=-1/6$ and using $\ln \mathbb{P}_N(\lambda_{\rm max}\le y, \alpha)=f(x)$
we find that 
\begin{equation}\label{eq:fxAsympt}
f(x)\approx - \frac{1}{16 \pi x^{\frac{3}{2}}}\,
 e^{-\frac{4 }{3} x^{\frac{3}{2}}}\;\; {\rm as}\;\; x\rightarrow \infty
\end{equation}
Hence $u^2(x)=-f''(x)\approx e^{-\frac{4 }{3} x^{\frac{3}{2}}}\: \frac{1}{4 \pi \sqrt{x}}$
and consequently as $x\to \infty$
\begin{equation}
\label{uxbc} 
u(x)=\sqrt{-f''(x)}\approx e^{-\frac{2 }{3} x^{\frac{3}{2}}}\: \frac{1}{2 \sqrt{\pi} x^{1/4}}.
\end{equation}

Finally integrating $f''(x)$ twice and using the appropriate boundary condition as 
$x\to \infty$, we get
\begin{equation}
f(x) = -\int_x^{\infty} ds (s-x) u^2(s)
\label{fxfinal1}
\end{equation}
where $u(x)$ satisfies the Painlev\'e II equation \eqref{ux} with the boundary condition
\eqref{uxbc}. Comparing to \eqref{eq:defTW}, we have thus shown that
the scaling function 
$f(x)=\ln F_2(x)$ where $F_2(x)$ is the Tracy-Widom function ($\beta=2$). This then
constitutes our derivation for the Tracy-Widom distribution for the GUE ($\beta=2$).
\\

Using recursion relations that we have derived for orthogonal polynomials
on a semi-infinite interval, we have shown that the large $N$
asymptotics of the distribution of the maximal eigenvalue
of a Gaussian random matrix (from the GUE)
is described in the double scaling regime by the Painlev\'e II equation.
Similar recursion relations for other orthogonal polynomials
leading to different kinds of Painlev\'e equations
have also been established in a number of papers, see
\cite{forresterbis} and references therein, in particular 
\cite{chen,periwal,Marino}.

\section{Conclusion}

In this paper, 
we have provided a rather simple and pedestrian derivation of the Tracy-Widom
law for the distribution of the largest eigenvalue of a Gaussian unitary random matrix.
This was done by suitably adapting a method of orthogonal polynomials developed by 
Gross and Matytsin~\cite{gross} in the context of two dimensional Yang-Mills theory. 
Our derivation requires just elementary asymptotic scaling analysis of a pair of coupled 
nonlinear recursion
relations. Strictly in the $N\to \infty$ limit, there is a $3$-rd order phase transition
in the form of the probability distribution of $\lambda_{\rm max}$ as $\lambda_{\rm max}$ crosses
its mean value from left to right. For finite but large $N$, the two regions are connected by
a smooth crossover function and the shape of this crossover function is precisely
the Tracy-Widom distribution that describes the `typical' small fluctuations of
$\lambda_{\rm max}$ around its mean. The `atypical' large fluctuations to the left
and right of the mean are described by large deviation tails that correspond to
the two `phases' across this phase transition. In qualitative analogy
to the two-dimensional Yang-Mills theory, the left (left large deviation)
and the right (right large deviation) phases correspond respectively to the `strong'
and `weak' coupling phases of the two-dimensional QCD.     
Apart from the simple derivation of the Tracy-Widom GUE law, 
we were also able to compute
the precise right large deviation tail of the maximal eigenvalue distribution that
is not described by the Tracy-Widom distribution. In the language of QCD, this
right tail corrections are similar 
to the non-perturbative (in $1/N$ expansion) corrections
to the QCD partition function in $2$-d~\cite{gross}. 

One drawback of our method is that it works only for the GUE case (with Dyson index $\beta=2$). It would be
challenging to see if this method can be extended/generalized to derive the Tracy-Widom law for
the other two Gaussian ensembles, namely the GOE ($\beta=1$) and the GSE ($\beta=4$).  
In addition, this method for $\beta=2$ should be useful to compute the largest eigenvalue
distribution for other non-Gaussian matrix ensembles, such as the Laguerre (Wishart
matrices) or the Jacobi ensembles.

\end{document}